\documentclass[12pt]{article}
\usepackage{amsmath, amsthm,comment} 
\usepackage{amsfonts}
\usepackage{amssymb,graphics,psfrag}
\usepackage{array,epsfig,stmaryrd,graphicx}

\def\hybrid{\topmargin -20pt    \oddsidemargin 0pt
        \headheight 0pt \headsep 0pt
        \textwidth 6.25in       
        \textheight 9.5in       
        \marginparwidth .875in
        \parskip 5pt plus 1pt   \jot = 1.5ex}
\hybrid
\numberwithin{equation}{section}
\numberwithin{table}{section}
\numberwithin{figure}{section}
\def\a{\alpha}
\def\b{\beta}

\def\t{\tau}
\def\l{\lambda}

\def\th{\vartheta}

\newcommand{\beq}{\begin{equation}}
\newcommand{\eeq}{\end{equation}}
\newcommand{\be}{\begin{equation}}
\newcommand{\ee}{\end{equation}}
\newcommand{\bea}{\begin{eqnarray}}
\newcommand{\eea}{\end{eqnarray}}   
\newcommand{\ben}{\begin{eqnarray*}}
\newcommand{\een}{\end{eqnarray*}}                  
\newcommand{\ba}{\begin{aligned}}
\newcommand{\ea}{\end{aligned}}
\newcommand{\bt}{\begin{tabular}}
\newcommand{\et}{\end{tabular}}
\newcommand{\bc}{\begin{center}}
\newcommand{\ec}{\end{center}}

\newcommand{\mbb}{\mathbb}
\newcommand{\mc}[1]{\mathcal{#1}}

\newcommand{\RE}{\textrm{Re} \,}

\newcommand{\cref}{{\bf [check ref]}}

\newcommand{\N}{\Theta}

\newcommand{\bs}{\begin{subarray}{c}}
\newcommand{\es}{\end{subarray}}

\newcommand{\CF}{{\cal F}}

\newcommand{\CN}{{\cal N}}

\newcommand{\CP}{{\cal P}}

\newcommand{\re}{{\rm e}}
\newcommand{\ri}{{\rm i}}

\newcommand{\Li}{{\rm Li}}

\newcommand{\vt}{\vartheta}
\newcommand{\ph}{\phantom}
\psfrag{a1}{$\alpha_1$}
\psfrag{a2}{$\alpha_2$}
\psfrag{a3}{$\alpha_3$}
\psfrag{a4}{$\alpha_4$}
\psfrag{a5}{$\alpha_5$}
\psfrag{a6}{$\alpha_6$}
\psfrag{a7}{$\alpha_7$}
\psfrag{a8}{$\alpha_8$}

\def\blfootnote{\xdef\@thefnmark{}\@footnotetext} 
\long\def\symbolfootnote[#1]#2{\begingroup%
\def\thefootnote{\fnsymbol{footnote}}\footnote[#1]{#2}\endgroup}
\begin{document}

\begin{titlepage}

\hfill\vbox{
\hbox{CERN-PH-TH/2007-084}
}

\vspace*{ 2cm}
\begin{center}
{\Large \bf  Topological amplitudes in heterotic strings with Wilson lines} 
\end{center}
\medskip

\vspace*{4.0ex}

\centerline{\large \rm
Marlene Weiss$^a$}
\vspace*{1.8ex}
\begin{center}

{\em Department of Physics, CERN\\[.1cm]
 Geneva 23, CH-1211 Switzerland}

\vspace*{1ex}
{and}
\vspace*{1ex}

{\em ETH Zurich\\[.1cm]
CH-8093 Z\"urich, Switzerland}

\symbolfootnote[0]{\tt $^a$marlene.weiss@cern.ch}

\vskip 0.5cm
\end{center}

\centerline{\bf Abstract}
We consider d=4, $\CN=2$ compactifications of heterotic strings with an arbitrary number of Wilson lines. In particular, we focus on known chains of candidate heterotic/type II duals. We give closed expressions for the topological amplitudes $F^{(g)}$ in terms of automorphic forms of $SO(2+k,2,\mbb{Z})$, and find agreement with the geometric data of the dual K3 fibrations wherever those are known.
\medskip

\vskip 1cm

\noindent
\end{titlepage}
\tableofcontents
\section{Introduction}
Over the last decade, tremendous progress has been made in establishing and understanding $d=4,\ \CN=2$ heterotic-type II duality, which connects the heterotic string compactified on $K3\times \mbb{T}^2$ with compactifications of type IIA theory on $K3$-fibrations. One of the most fruitful approaches has been to compute the low energy effective action for models with explicitly known heterotic and type II realizations. More precisely, the 4d effective action of these $\CN=2$ compactifications has been known for a long time to contain a series of BPS protected higher-loop terms of the form
\be
S\sim\int F^{(g)}(t,\bar{t})T^{2g-2}R^{2}+\cdots,
\ee
where $R$ is the Riemann tensor, $T$ the graviphoton field strength, and the couplings $F^{(g)}$ are amplitudes of the topological string on the internal Calabi-Yau \cite{agnt1,bcov}. On the heterotic side, these amplitudes appear at 1--loop \cite{agnt2} and are therefore in general accessible to computation \cite{harvey-moore,mm2,kkrs,km}. The result can be mapped to the type II side, yielding striking predictions in enumerative geometry.\\
The amplitudes $F^{(g)}$ are also intriguing from a mathematical point of view, as they involve interesting classes of automorphic functions. Furthermore, the Higgs transitions on the heterotic side correspond to geometric transitions between the corresponding Calabi-Yaus on the type II side. A more precise picture of how the heterotic moduli spaces are connected might therefore provide some insight into the web of type II vacua.\\
Until now, most explicit comparisons between heterotic and type II models have been restricted to cases with a small number $n_v$ of massless Abelian vector multiplets, namely $n_v=3,4,5$. These vector multiplets are the graviphoton, the heterotic dilaton $S$, one or two ($n_v=4$) moduli $T,U$ from the compactification torus, and if $n_v=5$, one Wilson line modulus $V$. However, by now there is a myriad of conjectured heterotic-type II pairs with higher numbers of vector multiplets waiting to be analyzed.\\
In \cite{afiq}, the authors obtained chains of heterotic-type II duals by compactifying the heterotic string on $K3\times \mbb{T}^2$ in various orbifold realizations. In each chain, subsequent models are connected by a sequential Higgs mechanism reducing the number of generic Wilson line moduli by one. $K3$ is realized as an orbifold $\mbb{T}^4/\mbb{Z}_N$, $N=2,3,4,6$ and the $\mbb{Z}_N$ is simultaneously embedded in the gauge connection in a modular invariant way. For the last models in the chains, the candidate type II duals can be explicitly constructed.\\
The classical vector multiplet moduli space of compactifications with $k=n_v-4$ Wilson lines is given by the special K\"ahler space
\be
{SU(1,1)\over U(1)}\times{SO(2+k,2)\over SO(2+k)\times SO(2)},
\ee
where the first factor corresponds to the dilaton and the second to the torus and Wilson line moduli. The T-duality group, under which the vector multiplet couplings have to transform as automorphic functions, is $SO(2+k,2;\mbb{Z})$ \cite{cdfv,dkll,ms}.\\
For the $SO(2,2;\mbb{Z})$ case with four vector multiplets, i.e. the well-known STU model, the higher derivative couplings have been computed in \cite{mm2}. They can be expressed in terms of expansion coefficients of ordinary modular forms. The case with five vector multiplets (one Wilson line) has been studied at the level of prepotential and $F^{(1)}$ in \cite{ccl}. This case is somewhat special, as the T-duality group is here $SO(3,2;\mbb{Z})\cong Sp(4,\mbb{Z})$ \cite{ms}, and the corresponding automorphic functions are given by Siegel modular forms \cite{ms}. The effective couplings can be expressed in terms of Jacobi forms of index one, yielding a prescription how to split off the part depending on the Wilson line modulus from the gauge lattice.\\
The generic case involves more general automorphic forms. However, we can define a splitting procedure analogous to the one in \cite{ccl}, and the split lattice sum can be explicitly expressed in terms of ordinary Jacobi Theta functions. Once this split is determined, we can use the technique of lattice reduction \cite{borcherds} to explicitly compute higher-derivative F-terms for heterotic $\CN=2$ compactifications with an arbitrary number of Wilson lines. The final result involves the q-expansion coefficients of the moduli independent Higgsed part of the lattice sum. Even though the computation is done at the orbifold point, the results are fully valid at generic points of $K3$ moduli space, since the couplings $F^{(g)}$ only depend on vector multiplets and therefore cannot mix with the $K3$ moduli, belonging to hypermultiplets.\\
While the formalism can be applied to almost any symmetric $\mbb{Z}_N$ orbifold limit of $K3$, we mainly focus on the dual pairs found in \cite{afiq}. We compute the corresponding topological amplitudes $F^{(g)}$ in closed form. For genus zero, our results agree with the numbers of rational curves found on the type II side wherever those are known \cite{bkkm}. The present computation extends previous work on threshold corrections for models with a single Wilson line \cite{kawai,ccl,stieberger}, and also provides a more explicit realization, extended to higher genus, of the general results of \cite{hm}.\\
This paper is organized as follows. In section \ref{sec:het}, we review heterotic compactifications with $\CN=2$ supersymmetry and the Higgs chains of \cite{afiq}. In section \ref{sec:coupl}, we explain how to compute partition sums and higher derivative F-terms in general heterotic orbifold setups. Section \ref{sec:wil} introduces the lattice splits in the presence of Wilson lines. A general expression for the amplitudes $F^{(g)}$ in the presence of Wilson lines is derived. In section \ref{sec:dual}, we use our results to extract geometric information on the dual Calabi-Yau manifold. This provides a highly nontrivial check of our computation in those cases where instanton numbers are known on the type II side. Section \ref{sec:concl} contains some concluding remarks and further directions of research. Appendix \ref{ap:thet} summarizes some facts about Jacobi and Riemann-Siegel theta functions, and appendix \ref{ap:latred} reviews the Borcherds-Harvey-Moore technique of lattice reduction. Finally, appendix \ref{ap:data} collects tables of instanton numbers for several models discussed in the text.

\section{Heterotic $\CN=2$ compactifications}\label{sec:het}
In this section, we briefly discuss the construction of heterotic $\CN=2$ compactifications and their matter spectrum. There are two main approaches to analyzing these models. Section \ref{ssec:CY} reviews the purely geometrical approach of \cite{kv}, while section \ref{ssec:CFT} reviews the exact CFT construction via orbifolds of \cite{afiq}. Even though the two approaches are completely equivalent, it proves very useful to keep the two in mind simultaneously, as sometimes one is more convenient, sometimes the other. Section \ref{ssec:chains} reviews how these compactifications fall into chains of models connected by a sequential Higgs mechanism \cite{afiq}.
\subsection{The Calabi-Yau approach}\label{ssec:CY}
Consider compactification of the heterotic string on $K3\times \mbb{T}^2$. 
In order to break the gauge group $\mc{G}=E_8\times E_8$ of the ten-dimensional heterotic string down to a subgroup G, one gives gauge fields on K3 an expectation value in H, where $G\times H$ is a maximal subgroup of $\mc{G}$. Geometrically, this corresponds to embedding a H-bundle V on K3. This bundle can be chosen to be the tangent bundle of K3, an $SU(2)$-bundle with instanton number $\int_{K3}c_2(V)=24$. This is the standard embedding, where the spin connection on K3 is equal to the gauge connection. More generally, one can embed several stable holomorphic SU(N)-bundles $V_a$, as long as the constraints from modular invariance 
\be\label{cons1}
\sum_a c_2(V_a)=24 \qquad c_1(V_a)=0
\ee
are satisfied. We will here only consider embeddings of one or two $SU(2)$-bundles on one respectively both $E_8$ and write their instanton numbers according to \eqref{cons1} as $(d_1,d_2)=(12+n,12-n)$.\\
The number of gauge neutral hypermultiplets is determined as follows \cite{kv}. There is a universal gravitational contribution of 20, and each of the $SU(N_a)$-bundles $V_a \rightarrow K3$ with $\int_{K3}c_2(V_a)=A$ has an extra $AN_a+1-N_a^2$ moduli, therefore we get additional $45$ moduli for one and $51$ for two embedded $SU(2)$ bundles. The rank of the gauge group is reduced by the rank of the embedded bundle, N-1. For the standard embedding, we thus find 65 hypermultiplets and an enhanced gauge group $E_7\times E_8$, the first model in the $\mbb{Z}_2$ chain in \cite{afiq}. The Cartan subalgebra of $E_7\times E_8$ contains 15 generators, and there is an extra $U(1)^4$ from the SUGRA multiplet and torus compactification, therefore this model has $n_v=19$ vector multiplets.\\ 

\subsection{Exact CFT construction via orbifolds}\label{ssec:CFT}
Rather than following the approach presented above, we will here realize the heterotic models following \cite{afiq} in the so-called exact CFT construction via orbifolds. In this approach, the K3 is realized as a $\mbb{Z}_N$ orbifold, while simultaneously the spin connection is embedded into the gauge degrees of freedom. We will mainly concentrate on the $\mbb{Z}_N$--embeddings given in table \ref{tab:orb}. The orbifold $\mbb{Z}_N$ twist $\theta$ acts on two of the four complex bosonic transverse coordinates as $\re^{\pm{2\pi\ri\over N}}$. Since we impose $\CN=2$ SUSY, N can only take on the values $2,3,4,6$ \cite{hm}. The action of $\theta$ on the gauge degrees of freedom is strongly restricted by worldsheet modular invariance. We implement it as a shift of the gauge lattice, writing for the torus and gauge lattice sum
\be
{\bf Z}^{18,2}[^a_b]=\sum_{p\in\Gamma^{18,2}+a\gamma}\re^{2\pi\ri b \gamma\cdot p}q^{|p_L|^2\over 2}\bar{q}^{|p_R|^2\over 2},
\ee
where $a,b\in\{1/N,\cdots (N-1)/N\}$.
The shift $\gamma\in\Gamma^{18,2}$ has to fulfill the modular invariance and level-matching constraints \cite{Polchinski}
\be
\sum_{i=1}^8\gamma_i=\sum_{i=9}^{16}\gamma_i=0\ {\rm mod}\  2
\ee 
and
\be
\gamma^2=2\ {\rm mod}\ 2N.
\ee
One then finds the possible inequivalent $\mbb{Z}_N$ orbifolds: There are 2 for $\mbb{Z}_2$, 5 for $\mbb{Z}_3$, 12 for $\mbb{Z}_4$ and 61 for $\mbb{Z}_6$ \cite{stieberger}. Note that in those cases where the same type of shift is modular invariant for different N, those models are equivalent as far as the topological amplitudes $F^{(g)}$ are concerned. The reason for this is that they are only distinguished by the specific orbifold realization of the K3-surface. Since the moduli of the K3 live in hypermultiplets which do not mix with the vector multiplets, the higher-derivative couplings should be identical for the different $\mbb{Z}_N$ embeddings. They can however differ if we turn on Wilson line moduli corresponding to the gauge groups only present in the orbifold limit \cite{hm}, as will be explained in section \ref{ssec:moduli}.\\ 
Some non-standard embeddings, along with their perturbative gauge group, are given in table \ref{tab:other}. These groups are easily read off from the simple root system for $E_8$ given below, table \ref{tab:E8sr}. The unbroken group is generated by the roots $\alpha_i$ invariant under the shift $\gamma$, i.e. fulfilling
\be\label{inv}
\re^{2\pi\ri\gamma\cdot \alpha_i\over N}=1.
\ee
\begin{table}[!h]
\begin{center}\begin{tabular}{|l|l|l|l|}
\hline
$\mbb{Z}_2$&$\gamma^1$=(1,-1,0,0,0,0,0,0);&&\\&$\quad \gamma^2$=(0,0,0,0,0,0,0,0)&$SU(2)\times E_7\times E_8'$&$n$=12\\ 
$\mbb{Z}_3$&$\gamma^1$=(1,1,2,0,0,0,0,0);&&\\&$\qquad \gamma^2$=(1,-1,0,0,0,0,0,0)&$SU(3)\times E_6 \times U(1)'\times E_7' $&$n$=6\\
$\mbb{Z}_4$&$\gamma^1$=(1,1,1,-3,0,0,0,0);&&\\&$\qquad \gamma^2$=(1,1,-2,0,0,0,0,0)&$SO(10)\times SU(4)\times E_6'\times SU(2)'\times U(1)'$&$n$=4\\
$\mbb{Z}_6$&$\gamma^1$=(1,1,1,1,-4,0,0,0);&&\\&$\qquad \gamma^2$=(1,1,1,1,1,-5,0,0)&$SU(5)\times SU(4)\times U(1)\times SU(6)'\times SU(3)'\times SU(2)'$&$n$=2\\
\hline
\end{tabular}\end{center}
\caption{Embeddings of the spin connection in the gauge degrees of freedom}\label{tab:orb}
\end{table}
\begin{table}[!h]
\begin{center}\begin{tabular}{|c c c c c c c c | c|}
\hline
0&1&1 &0 &0 &0 &0 &0 & $\alpha_1$\\
0& 0&-1& 1& 0& 0& 0& 0&  $\alpha_2 $\\
0& 0& 0&-1 & 1 & 0 &0 & 0&  $\alpha_3 $\\
0& 0& 0& 0 & -1& 1 &0 & 0 &  $\alpha_4 $\\
0& 0& 0& 0 & 0 &-1 & -1 & 0 &  $\alpha_5 $\\
0& 0& 0& 0 & 0 & 0 & 1 & 1 &  $\alpha_6 $\\
-${1\over 2} $& -${1\over 2} $& ${1\over 2} $& ${1\over 2} $& ${1\over 2} $& ${1\over 2} $& -${1\over 2} $& -${1\over 2} $& $ \alpha_7 $\\
0& 0& 0& 0 & 0 & 0 & 1 & -1 &  $\alpha_8 $\\
&&&&&&&&\\
\hline
\end{tabular}\end{center}
\caption{A simple root system for $E_8$}\label{tab:E8sr}
\end{table}
\begin{figure}
\centering
\be
\left(\begin{array}{cccccccc}
2 & -1 & 0&&\cdots&&&0\\
-1&2&-1&0&&&&0\\
0&-1&2&-1&0&&&0\\
\vdots &0&-1&2&-1&0&&0\\
&&0&-1&2&-1&0&-1\\
&&&0&-1&2&-1&0\\
&&&&0&-1&2&0\\
0&&\cdots&&-1&0&\cdots&2
\end{array}\right)
\ee
\caption{Cartan matrix of $E_8$}\label{fig:cartan}
\end{figure} 
In the first embedding in table \ref{tab:other}, the invariant roots on the first $E_8$ are the $126$ roots of $E_7$, generated by the roots $\alpha_2,\cdots,\alpha_8$. One realization is given in table \ref{tab:E8sr}. For a general $\mbb{Z}_N$ embedding, the gauge group from the first $E_8$ would then be $U(1)\times E_7$. For $N=2$, $\gamma$ itself is also a root, orthogonal to the others, fulfilling \eqref{inv}, and the $U(1)$ is enhanced to an $SU(2)$. On the second $E_8$, the invariant roots are the roots of $SO(14)$ $\alpha_1,\cdots,\alpha_6,\alpha_8$, and an extra root $(1,-1,0^6)$ such that the unbroken gauge group is $SO(16)$. The second embedding is obviously analogous, only in this case $N=3$, therefore $(1,-1,0^6)$ is not an invariant root anymore. For the left-hand side of the third embedding, the unbroken roots are $\alpha_1,(1,-1,0^6)$, and the second system, orthogonal to the first $\alpha_3,\cdots\alpha_8$, yielding a perturbative gauge group $SU(3)\times E_6$. On the second $E_8$, the unbroken roots are $\alpha_1,\cdots,\alpha_7,({1\over 2},-{1\over 2},-{1\over 2},-{1\over 2},-{1\over 2},-{1\over 2},{1\over 2},-{1\over 2})$, forming the Dynkin diagram of $SU(9)$. The other examples work out similarly.
 Note that each of these realizations breaks the original gauge group $E_8\times E_8$ to a different rank 16 subgroup, containing a nonabelian rank $r$ group $G$ and a $U(1)^{16-r}$ that may be enhanced as in the example above. However, this latter factor is only present in the orbifold limit; for a smooth K3, the gauge group consists merely of $G$.

The perturbative gauge group $G\times G'$ can subsequently be spontaneously broken to a subgroup $G_1\subset G$ via maximal Higgsing, as explained in section \ref{ssec:CY} within the Calabi-Yau approach of \cite{kv}. This subgroup depends on the embedding $\gamma$ only via its instanton numbers: For the standard embedding with $n=12$, there are no instantons on the second $E_8$ and the gauge group $E_8'$ can not be broken at all. For the cases $n=0,1,2$, complete Higgsing is possible. For $n=3,4,6,8$, there are too few hypermultiplets on $E_8'$ that could be used for Higgsing, and $G'$ can only be broken to a terminal subgroup $G_1=SU(3), SO(8),E_6,E_7$ \cite{stieberger}. Once again, we consider the standard $\mbb{Z}_2$ orbifold as an example. The hypermultiplets in the untwisted ($\theta^0$) and twisted ($\theta^1$) sectors transform under $E_7\times SU(2)$ in the following representations:
\be
\ba
(56,2)+4(1,1) \qquad& ({\rm untwisted},\ \theta^0)\\ 
8\left((56,1)+4(1,2)\right)\qquad & ({\rm twisted},\ \theta^1).
\ea
\ee
 We can now Higgs the $SU(2)$ giving vevs to three scalars, and we are left with 10 hypermultiplets transforming in the $\mathbf{56}$ of $E_7$ and 65 singlet hypermultiplets, as advertised in section \ref{ssec:CY}. We can then break $E_7$ further by sequential Higgs mechanism. Since the instanton numbers corresponding to this embedding are $(24,0)$, we can not break the $E_8'$ from the second $E_8$ lattice at all. A complete classification of orbifold limits of $K3$ along with their instanton numbers can be found in \cite{stieberger}.   

\begin{table}[!h]\label{tab:other}
\begin{center}\begin{tabular}{|l|l|l|l|}
\hline
$\mbb{Z}_2$&$(1,-1, 0, 0, 0, 0, 0, 0);$&&\\&$\qquad (2, 0,0,0,0,0,0,0)$& $SU(2)\times E_7\times SO(16)'$&$n=4$\\ 
$\mbb{Z}_3$&$(2, 0, 0, 0, 0, 0, 0, 0);$&&\\&$\qquad( 2, 0,0,0,0,0,0,0)$&$U(1)\times SO(14)\times U(1)'\times SO(14)'$&$n=0$\\
$\mbb{Z}_3$&$(1, 1,-2, 0, 0, 0, 0, 0);$&&\\&$\qquad(-2, 1,1,1,1,1,2,1)$&$SU(3)\times E_6\times SU(9)'$&$n=3$\\
$\mbb{Z}_4$&$(3,-1, 0, 0, 0, 0, 0, 0);$&&\\&$\qquad( 0, 0,0,0,0,0,0,0)$ & $SU(2)\times U(1)\times SO(12)\times E_8'$&$n=12$\\
$\mbb{Z}_6$&$(3,-1,-1,-1,-1,-1, 1, 1);$&&\\&$\qquad( 3,-3,2,0,0,0,0,0)$& $U(1)^2\times SU(7)\times U(1)'\times SU(2)'^2\times SO(10)'$&$n=2$\\
\hline
\end{tabular}\end{center}
\caption{Other $\mbb{Z}_N$ embeddings of the spin connection}
\end{table}

\subsection{Chains of dual models and the sequential Higgs mechanism}\label{ssec:chains}
Once one has chosen a modular invariant embedding of $SU(N)$ bundles, and maximally Higgsed the gauge group on the $E_8$ lattice where the embedding has the lower instanton number, one can perform a cascade breaking on the remaining gauge group along the chain $E_8\rightarrow E_7\rightarrow E_6 \rightarrow SO(10)\rightarrow SU(5)\rightarrow SU(4)\rightarrow SU(3)\rightarrow SU(2)\rightarrow {\rm (nothing)}$. For the example of the standard $\mbb{Z}_2$ orbifold, this goes as follows.\\
Starting with the (65,19) model with $E_7\times E_8$ symmetry remaining after the gauge embedding, one can move to a point in moduli space where the $E_7$ gauge symmetry is restored. Under the maximal subgroup $E_6\times U(1)\in E_7$, the $\bf{56}$ of $E_7$ decomposes as $\bf{56}=\bf{27}+\overline{\bf{27}}+\bf{1}+\bf{1}$. At this point, there are 10 $\bf{56}$, therefore 20 $E_6$ singlets charged under the U(1). We now give a generic vev to the adjoint scalars in the unbroken vector multiplets, thereby giving masses to all hypermultiplets charged with respect to $E_6$, and at the same time breaking $E_6$ to its maximal Abelian subgroup $U(1)^6$. Using one scalar to Higgs the $U(1)$, we get 19 extra gauge singlet fields: the new spectrum is $(84,18)$, the second model in the corresponding chain in \cite{afiq}. We can then move to a point in moduli space where the $U(1)^6$ is enhanced to $E_6$ and continue this procedure until no gauge symmetry remains on this lattice. In this way, one easily finds a chain of models with characteristics $(n_h,n_v)$ \cite{afiq} 
\be
(65,19),(84,18),(101,17),(116,16),(167,15),(230,14),(319,13),(492,12)
\ee
The same mechanism can be applied to the other embeddings in table \ref{tab:orb}. For the $\mbb{Z}_3$ orbifold, $n=6$, therefore we can maximally Higgs on the second lattice down to $E_6$. On the first $E_8$ lattice, we first Higgs down to the rank-reduced subgroup and then start cascade breaking as explained above. The result is a chain $E_6\rightarrow SO(10)\rightarrow \cdots\rightarrow SU(2)\rightarrow 0$ passing through models with characteristics
\be
(76,16),(87,15),(96,14),(129,13),(168,12),(221,11),(322,10).
\ee
For the $\mbb{Z}_4$ orbifold, $n=4$, maximal Higgsing leaves an $SO(8)$ on the second lattice and the embedding of the spin connection leaves a rank-reduced subgroup $SU(4)$ on the first. The resulting chain reads
\be
(123,11),(154,10),(195,9),(272,8).
\ee
The $\mbb{Z}_6$ orbifold in table \ref{tab:orb}, finally, has $n=2$ and therefore allows for complete Higgsing. The rank-reduced subgroup is $SU(5)$, Higgsed via the chain
\be
(118,8),(139,7),(162,6),(191,5),(244,4).
\ee
The last four models in each chain have candidate type II duals, i.e. known K3 fibrations with the right Betti numbers. It is interesting to note that on the type-II side, the cascade breaking procedure corresponds precisely to moving between moduli spaces of different Calabi-Yau manifolds.  
Indeed, as pointed out in \cite{kv}, this is strikingly similar to the specific type-II process described in \cite{gms}.
\section{Higher derivative couplings for $\mbb{Z}_n$ orbifolds}\label{sec:coupl}

We will consider here the $E_8\times E_8$ formulation of the 10 dimensional heterotic string, where the gauge degrees of freedom are encoded by 16 left-moving bosons, and compactify it on $K3\times \mbb{T}^2$, yielding another two left- and two right-moving bosons. These fields take their values on an even self--dual lattice of signature $(18,2)$ that will be denoted by $\Gamma^{18,2}$. One can identify $\Gamma^{18,2}$ as obtained from a Euclidean standard lattice by an $SO(18,2)$ rotation. The moduli space of inequivalent lattices is therefore given by
\be
{SO(18,2)\over SO(18)\times SO(2)}.
\ee
This homogeneous space can be parametrized following \cite{harvey-moore},\cite{hm} by
\be
u(y)=(\vec{y},y^+,y^-;1,-{1\over 2}(y,y)),\ y\in \mbb{C}^{17,1}
\ee
with $y_2>0,(y_2,y_2)<0$ and inner product
\be
(x,y)=(\vec{x},\vec{y})-x^+y^--x^-y^+.
\ee
The right-moving components of a vector in $\Gamma^{18,2}$ with respect to a vector $(\vec{b},m_-,n_+,m_0,n_0)$ in the fixed Euclidean standard lattice are then denoted by $p_R=p\cdot u(y)$, and we have
\be
{p_L^2-p_R^2\over 2}={1\over 2(y_2,y_2)}\bigl(\vec{b}\cdot \vec{b} + m_- n_+ +m_0 n_0 \bigr),
\ee
\be\label{param}
{p_R^2\over 2}={-1\over 2(y_2,y_2)}\bigl|\vec{b}\cdot\vec{y}+ m_+y^--n_-y^++n_0+{1\over 2}m_0(y,y)\bigr|^2,
\ee
The general expression for $F^{(g)}$ is given by \cite{agnt2,mos,km}
\be
\label{Fg}
F^{(g)}={1\over Y^{g-1}}\int_{\cal F} {d^2 \tau  \over \tau_2} {1\over |\eta|^4}\sum_{\rm even} {i\over \pi} \partial_{\tau}
\biggl( {\vartheta[^\a_\b](\tau) \over \eta(\tau)}\biggr) Z_g^{\rm int}[^\a_\b],
\ee
where
\be
Z_g^{\rm int}[^\a_\b]=\langle :\bigl( {\partial X}\bigr)^{2g}: \rangle=\CP_{g}C^{\rm int}_g[^\a_\b].
\ee
$\CP_{g}(q)$ is a one-loop correlation function of the bosonic fields and is given by \cite{lnsw},\cite{agnt2}
\be
\label{defphatg}
\re^{-\pi \lambda^2 \tau_2} \biggl( { 2\pi  \eta^3 \lambda \over \vartheta_1(\lambda|\tau)}\biggr)^2=
\sum_{g=0}^{\infty} (2 \pi \lambda)^{2g} {\cal P}_{g}(q),
\ee
and $C^{\rm int}_g[^a_b]$ denotes the trace over  the $(a,b)$ sector of the internal CFT with an insertion of $p_R^{2g-2}$, namely
\be
\sum_{a,b}c(a,b)(-1)^{2\a+2\b+4\a\b}{\vt[^\a_\b]\vt[^{\a+a}_{\b+b}]\vt[^{\a-a}_{\b-b}]\over \eta^3}\cdot Z_{4,4}[^a_b]\cdot Z^g_{\mbb{T}^2}[^a_b],
\ee
where $c(a,b)$ are constants ensuring modular invariance. \\
Note that for g=1, \eqref{Fg} is just the unregularized one-loop gravitational threshold correction
\be\label{Delta}
F^{(1)}=\int_{\cal F} {d^2 \tau  \over \tau_2^2} \biggl({\tau_2\over |\eta|^4}\sum_{\rm even} {i\over \pi} (-1)^{2\a +2\b + 4\a\b} \partial_{\tau}
\biggl( {\vartheta[^\a_\b](\tau) \over \eta(\tau)}\biggr){\widehat{E}_2\over 12} C^{\rm int}_g[^\a_\b]\biggr).
\ee

The contribution from the bosonic (4,4) blocks reads
\be
Z_{4,4}[^a_b]=16{\eta^2\bar{\eta}^2\over \vt^2[^{1-a}_{1-b}]\bar{\vt}^2[^{1-a}_{1-b}]}\qquad (a,b)\neq (0,0)
\ee
while the bosons on the $\mbb{T}^2$ together with the 16 bosons corresponding to the gauge degrees of freedom contribute \cite{hm}
\be
Z^g_{\mbb{T}^2}[^a_b]={1\over \eta^{18}}e^{-2\pi i ab\gamma^2}\sum_{p\in \Gamma^{18,2}+a\gamma}p_R^{2g-2}e^{2\pi i b\gamma\cdot p}q^{|p_L|^2\over 2}\bar{q}^{|p_R|^2\over 2}.
\ee
Using
\be
{\ri\over 4\pi}\sum_{\rm (\a,\b) even}(-1)^{2\a+2\a+4\a\b}\partial_\tau\left({\vt[^\a_\b]\over\eta}\right){\vt[^\a_\b]\vt[^{\a+a}_{\b+b}]\vt[^{\a-a}_{\b-b}]\over\eta^3}{Z_{4,4}[^a_b]\over |\eta|^4}=4{\eta^2\over \bar{\vt}[^{1+a}_{1+b}]\bar{\vt}[^{1-a}_{1-b}]},
\ee
one can write for \eqref{Fg}
\be\label{int}
F^{(g)}={1\over Y^{g-1}}\int_{\cal F} {d^2 \tau  \over \tau_2^2} \tau_2^{2g-1}\mathcal{P}_{2g}(q)\sum_{a,b}{c(a,b)\re^{2\pi\ri ab(2-\gamma^2)}\over \eta^{18}\vt[^{1+a}_{1+b}]\vt[^{1-a}_{1-b}]}\sum_{p\in \Gamma^{18,2}+a\gamma}p_R^{2g-2}e^{2\pi i b\gamma\cdot p}q^{|p_L|^2\over 2}\bar{q}^{|p_R|^2\over 2}.
\ee
The constants $c(a,b)$ can be determined by the modular invariance constraints \cite{hm}
\be
\ba
c(0,b)&=4\sin^4(\pi b)\\
c(a,b)&=\re^{\pi\ri a^2(2-\gamma^2)}c(a,a+b)\\
c(a,b)&=\re^{-2\pi\ri ab(2-\gamma^2)}c(b,-a).
\ea
\ee
Introducing the Siegel-Narain theta function with insertion and shifts (see Appendix \ref{ap:thet})
\be
\Theta^g_\Gamma(\tau,\gamma,a,b)=\sum_{p\in\Gamma+a\gamma}p_R^{2g-2}q^{|p_L|^2\over 2}\bar{q}^{|p_R|^2\over 2}\re^{\pi\ri b \gamma\cdot p},
\ee
we can rewrite \eqref{int} as
\be\label{int2}
F^{(g)}={1\over Y^{g-1}}\int_{\cal F} {d^2 \tau  \over \tau_2^2} \tau_2^{2g-1}\mathcal{P}_{2g}(q)\sum_{a,b}{c(a,b)\re^{2\pi\ri ab(2-\gamma^2)}\over \eta^{18}\vt[^{1+a}_{1+b}]\vt[^{1-a}_{1-b}]}\Theta^g_{\Gamma^{18,2}}(\tau,\gamma,a,b).
\ee
For the special cases of $\mathcal{N}$=2 compactifications with a factorized $\mbb{T}^2$, the prepotential and $F^{(1)}$ have been shown to be universal, i.e. independent of the specific model \cite{lnsw}.
In other words, they are identical for all compactifications on $K3 \times\mbb{T}^2$ with all Wilson lines set to zero. Everything then only depends on the torus moduli. It is easy to see that this also applies to the amplitudes $F^{(g)}$: When we set all Wilson line moduli to zero, the lattice sum obviously factorizes as
\be
\sum_{p\in\Gamma^{16,0}+a\gamma}q^{|p_L|^2\over 2}\re^{2\pi\ri bp\cdot\gamma}\sum_{\widehat{p}\in\Gamma^{2,2}} q^{|\widehat{p}_L|^2\over 2}\bar{q}^{|\widehat{p}_R|^2\over 2},
\ee  and we obtain
\be\label{int0}
\ba
F^{(g)}_{\rm 0 WL}=&{1\over Y^{g-1}}\int_{\cal F} {d^2 \tau  \over \tau_2^2} \tau_2^{2g-1}\mathcal{P}_{2g}(q)\sum_{a,b}{c(a,b)\over \eta^{18}\vt[^{1+a}_{1+b}]\vt[^{1-a}_{1-b}]}\sum_{p\in \Gamma^{16,0}+a\gamma}q^{p^2\over 2}\re^{\pi\ri g\gamma\cdot p}\Theta^g_{\Gamma^{2,2}}(\tau)\\
=&\int {d^2 \tau_2\over \tau_2^2} \tau_2^{2g-1}\mc{P}_{2g}\Theta^g_{\Gamma^{2,2}} {1\over \eta^{24}} \Omega,
\ea
\ee
where
\be
\Omega=\sum_{a,b}{c(a,b)\eta^6\over \vt[^{1+a}_{1+b}]\vt[^{1-a}_{1-b}]}\sum_{p\in \Gamma^{16,0}+a\gamma}q^{p^2\over 2}\re^{\pi\ri b\gamma\cdot p}.
\ee
For modular invariance, $\Omega$ then has to be a modular form of weight (10,0). Since the spaces of modular forms of even weight $2<w<12$ are one-dimensional, $\Omega$ has to be proportional to the single generator of weight 10 holomorphic modular forms $E_4E_6$. Indeed, one finds easily
 \be
\Omega=\sum_{a,b}{\eta^6\over \vt[^{1+a}_{1+b}]\vt[^{1-a}_{1-b}]}\sum_{A,B\\{\in \{0,1\}}}\prod_{i=1}^8\vt[^{A+a\gamma_i}_{B+b\gamma_i}]
\ee
which can be checked to be $-E_4E_6$. An abstract proof of this identity based on 6d anomaly cancellation can be found in \cite{kiritsis}.
We thus find that \eqref{int0} yields precisely the expression for the STU-model without Wilson line moduli given in \cite{mm2}. This universality property is related to the structure of the elliptic genus \cite{lnsw,lerche}. 
\\
We will now consider the nontrivial case with non-vanishing Wilson lines. The lattice sum does not factorize completely anymore. However, it should factorize partly, into a preserved and a Higgsed part. Indeed, it turns out that one can now write $F^{(g)}$ as
\be\label{int3}
F^{(g)}={1\over Y^{g-1}}\int_{\cal F} {d^2 \tau  \over \tau_2^2} \tau_2^{2g-2}\bar{\mathcal{P}}_{2g}(q)\sum_{a,b}{c(a,b)\re^{2\pi\ri ab(2-\gamma^2)}\over \eta^{18}\vt[^{1+a}_{1+b}]\vt[^{1-a}_{1-b}]}\sum_{J}\bar{\Theta}^g_{J,k}(\tau)\Phi_J^k[^a_b](q)
\ee
with
\be\label{thetaJ}
\bar{\Theta}^g_{J,k}=\sum_{p\in\Gamma^{k+2,2}_J}\bar{p}_R^{2g-2}q^{|p_L|^2\over 2}\bar{q}^{|p_R|^2\over 2},
\ee
where $\Gamma^{k+2,2}_J$ denotes the conjugacy class $J$ inside the lattice $\Gamma^{k+2,2}$, and $\Phi^k_J[^a_b](q)$ is a sum over theta functions that will be determined in the following section.
Note that \eqref{int3} is manifestly automorphic under the T-duality group $SO(2+k,2;\mbb{Z})$, since it has the structure of a Borcherds' type one-loop integral \cite{borcherds}.  
\section{Wilson lines: Splitting the lattice}\label{sec:wil}
\subsection{Decompositions of the $E_8$ lattice}\label{ssec:split}
Recall from section \ref{ssec:chains} that the sequential Higgs mechanism is realized by moving along specific branches of moduli space, away from the generic point. This corresponds to imposing constraints on the Wilson line moduli, such that at each step in the chain, the number of free Wilson line moduli is reduced by one. The lattice then splits non-trivially into a Higgsed part with $p\cdot y=0$ and a part depending on the remaining unconstrained moduli from Wilson lines and the torus.

\begin{figure}[!h]
\centering
\includegraphics[scale=.4]{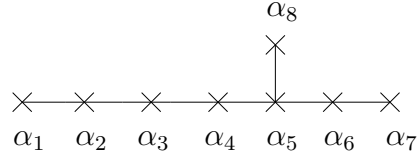}
\caption{$E_8$ Higgsed completely (no Wilson lines)}\label{fig:E8}
\end{figure}

First of all, we will determine how the lattice sum of $E_8$ behaves under decomposition into the maximal subgroups involved in the cascade breaking.
Consider the Dynkin diagram of $E_8$ (Fig. \ref{fig:E8}) and the simple root system given in table \ref{tab:E8sr}. In all the figures, crosses correspond to Higgsed generators of the group, while the generators remaining in the Coulomb phase due to Wilson lines are shown as circles.
Note that as can be seen from the labeling of the Dynkin diagram, the subgroup $E_7$ of $E_8$ is spanned by $\alpha_2,\cdots,\alpha_8$, $E_6$ by $\alpha_3,\cdots,\alpha_8$, $E_5=SO(10)$ by $\alpha_4,\cdots,\alpha_8$, and so on for $SU(5),SU(4),SU(3),SU(2)$. We denote the simple roots of the second $E_8$ by $\alpha'_i$.

We can now turn on one Wilson line, $y\sim \alpha_1$. On the other hand, turning on seven Wilson line moduli can be encoded in the constraint $\alpha_1\cdot y=0$. Both cases result in a split of the lattice sum of $E_8$ into 
\be\label{1wllat}
\ba
\sum_{p\in \Gamma_{E_8}}q^{p^2\over 2}&=\sum_{n_i\in\mbb{Z}}q^{n_1^2+\cdots+n_8^2-n_1n_2-n_2n_3-n_3n_4-n_4n_5-n_5n_6-n_5n_8-n_6n_7}\\
&=\sum_{n_i\in\mbb{Z}}q^{(n_1-{n_2\over 2})^2+{3\over 4}n_2^2+n_3^2+\cdots+n_8^2-n_2n_3-\cdots-n_6n_7}\\
&=\sum_{j=0,1}\sum_{n_1}q^{(n_1-{j\over 2})^2}\sum_{n_2,\cdots n_8\in\mbb{Z}}q^{{3\over 4}(2n_2-j)^2+n_3^2+\cdots+n_8^2-(2n_2-j)n_3-\cdots-n_6n_7}\\
&=\sum_{j=0,1}\vt[^{j/2}_{\phantom{j}0}](2\cdot)\sum_{n_2,\cdots,n_8}q^{{3\over 4}(2n_2-j)^2+n_3^2\cdots +n_8^2-(2n_2-j)n_3-\cdots-n_6n_7}.\\
\ea
\ee
Here and in the following, arguments $(m\cdot)$ stand for $m\cdot\tau$, see appendix \ref{ap:thet}.
The second sum in the last line is nothing else than the sum over the conjugacy class of $E_7$ corresponding to $(\alpha_1,p)=j$:
\be
\ba
(\alpha_1,p)&=2n_1-n_2\stackrel{!}{=}j\qquad \Rightarrow n_2=2n_1-j\\
&\Rightarrow p=n_1\alpha_1+(2n_1-j)\alpha_2+n_3\alpha_3+\cdots+n_8\alpha_8,\\
& p^2={3\over 2}(2n_1-j)^2+{j^2\over 2}+2n_3^2-2n_3(2n_1-j)-\cdots
\ea
\ee
and therefore
\be
q^{j^2\over 4}\sum_{n_2,\cdots,n_8}q^{{3\over 4}(2n_2-j)^2+n_3^2\cdots +n_8^2-(2n_2-j)n_3-\cdots-n_7n_8}=\sum_{(p,\alpha_1)=j}^{E_8}q^{p^2\over 2}=q^{j^2\over 4}\sum_{E_7^{(1)}}q^{p^2\over 2}.
\ee
We can also express the above in terms of theta functions. Rewriting the exponent in the second sum in the last line of \eqref{1wllat} as a sum over $p$ with $(p,\alpha_1)=0$ i.e. as
\be
\ba
p&=(n_1-{j\over 2})\alpha_1+(2n_1-j)\alpha_2+n_3\alpha_3+\cdots n_8\alpha_8\\
&=(-{n_7\over 2},n_1-{j\over 2}-{n_7\over 2},-n_1+{j\over 2}+{n_7\over 2},2n_1-j-n_3+{n_7\over 2},n_3-n_4+{n_7\over 2},\\
&\hspace{1.5cm} n_4-n_5+{n_7\over 2},-n_5+n_6-{n_7\over 2}+n_8,n_6-{n_7\over 2}-n_8),
\ea
\ee
we can write this sum as
\be
\ba
&\sum_{n_2,\cdots,n_8}q^{{3\over 4}(2n_2-j)^2+n_3^2\cdots +n_8^2-(2n_2-j)n_3-\cdots-n_7n_8}=\sum_{p\in E_7^{(1)}}q^{p^2\over 2}=\sum_{\bs p\in\Gamma_{E_8}-j{\alpha_1\over 2}\\(p,\alpha_1)=0\es} q^{p^2\over 2}\\
&=\sum_{\bs N_1,N_3,\cdots N_8\\N_3+\cdots+N_8=j\ {\rm mod}\ 2\\a=0,1\es}q^{(N_1-{j\over 2}-{a\over 2})^2}q^{{1\over 2}\left((N_3-{a\over 2})^2+\cdots+(N_8-{a\over 2})^2\right)}\\
&=\sum_{\bs N_1,\cdots N_8\in \mbb{Z}\\ a=0,1\\b=0,1\es}q^{(N_1-{j\over 2}-{a\over 2})^2}q^{{1\over 2}\left((N_3-{a\over 2})^2+\cdots+(N_8-{a\over 2})^2\right)}(-1)^{b(N_3+\cdots+N_8-j)}\\
&=\sum_{a,b\in\{0,1\}}\th[^{a/2+j/2}_{\phantom{a/2}0}](2\cdot)\th[^{a/2}_{b/2}]^6(-1)^{jb}.
\ea
\ee

\begin{figure}[htb]
\centering
\includegraphics[scale=.4]{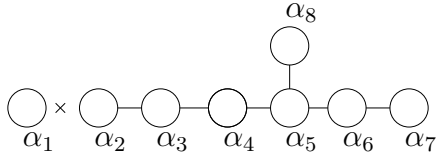}
\caption{$E_8\rightarrow E_7\times SU(2)$ }\label{fig:E81}
\end{figure}

We thus have decomposed the $E_8$-lattice according to $P_{E_8}\rightarrow P_{E_7^{(0)}}P_{A_1^{(0)}}+P_{E_7^{(1)}}P_{A_1^{(1)}}$, as shown in figure \ref{fig:E81}. This split has already been constructed in \cite{ccl}. Indeed \eqref{1wllat} is completely equivalent to the hatting procedure for Jacobi theta functions developed in \cite{ccl} for this particular split.

\begin{figure}[htb]
\centering
\includegraphics[scale=.4]{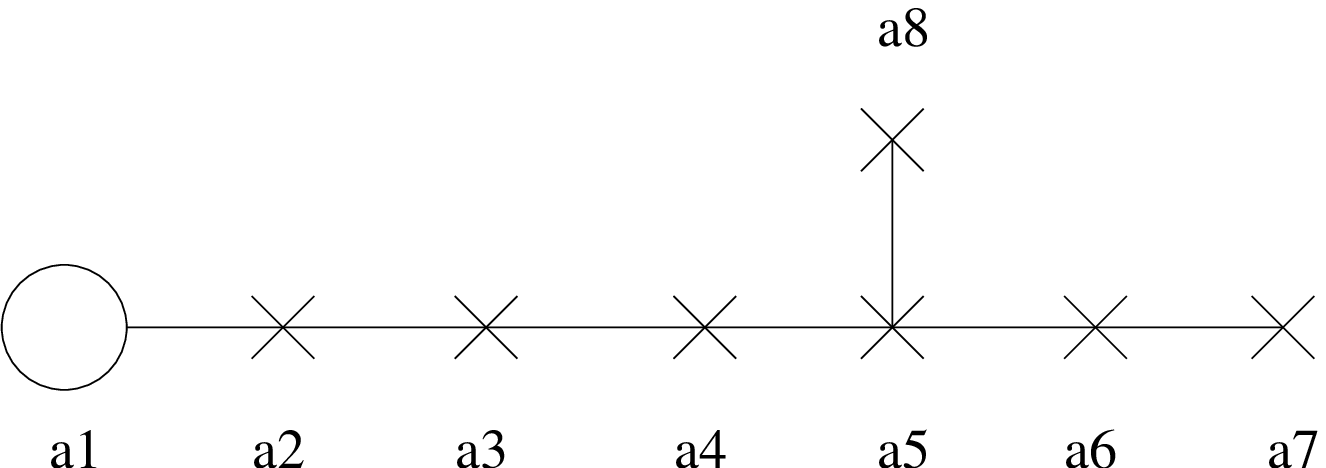}
\caption{$E_8$ with 1 Wilson line}\label{fig:E81W}
\end{figure}
The same procedure applies when we split the lattice in other maximal subgroups. Namely, we can decompose with respect to $E_8\supset E_6\times SU(3)$ :
\be\label{2wllat}
\ba
\sum_{p\in \Gamma_{E_8}}q^{p^2\over 2}&=\sum_{j_2=0,1,2}\sum_{\begin{subarray}{c}n_1,n_2\in\mbb{Z}\\j_1\in\{0,1\}\end{subarray}}q^{(n_1-{j_1\over 2})^2+3(n_2+{j_1\over2}-{j_2\over 3})^2}\sum_{n_3,\cdots,n_8\in\mbb{Z}}q^{{2\over 3}(3n_3-j_2)^2+n_4^2+\cdots+n_8^2-(3n_3-j_2)n_4-\cdots -n_6n_7}\\
&=\sum_{\bs j_1=0,1\\j_2=0,1,2\es}\th[^{j_1/2}_{\ph{a}0}](2\cdot)\th[^{
j_1/2+j_2/3}_{\ph{a/2}0}](6\cdot)\sum_{a,b\in\{0,1\}}\th[^{a/2+j_2/3}_{\ph{a}b/2}](3\cdot)\th[^{a/2}_{b/2}]^5(-1)^{b\cdot j_2}\\
&=P_{E_6^{(0)}}\cdot P_{A_2^{(0)}}+2 P_{E_6^{(1)}}\cdot P_{A_2^{(1)}},
\ea
\ee
The last relation in \eqref{2wllat} follows from
\be
\sum_{n_3,\cdots,n_8\in\mbb{Z}}q^{6(n_3-{j\over3})^2+n_4^2+\cdots+n_8^2-n_3n_4-\cdots -n_6n_7}
=q^{-{j^2\over 3}}\sum_{\begin{subarray}{c}p\in\Gamma_{E_8}\\(p,\alpha_1)=0\\(p,\alpha_2)=j\end{subarray}}q^{p^2\over 2}
=\sum_{E_6^{(j)}}q^{p^2\over 2},
\ee
and from the fact that $E_6^{(j=1)}$ and $E_6^{j=2}$ are equivalent. This case corresponds to 2 respectively 6 Wilson lines.\\
Analogously, we have lattice decompositions with respect to $E_8\supset SO(10)\times SU(4)$ (3 or 5 Wilson lines)

\be\label{3wllat}
\ba
&\sum_{p\in \Gamma_{E_8}}q^{p^2\over 2}\\
&=\sum_{j_3=0,1,2,3}\sum_{\begin{subarray}{c}n_1,n_2,n_3\in\mbb{Z}\\j_1\in\{0,1\}\\j_2\in\{0,1,2\}\end{subarray}}q^{(n_1-{j_1\over 2})^2+3(n_2+{j_1\over2}-{j_2\over 3})^2+6(n_3+{j_2\over 3}-{j_3\over 4})^2}\sum_{n_4,\cdots,n_8\in\mbb{Z}}q^{{3\over 8}(4n_4-j_3)^2+\cdots+n_8^2-(4n_4-j_3)n_5-\cdots -n_6n_7}\\
&=\sum_{j_3=0,1,2,3}\sum_{\begin{subarray}{c}j_1=0,1\\j_2=0,1,2\end{subarray}}\th[^{j_1/2}_{\ph{a}0}](2\cdot)\th[^{
j_2/3-j_1/2}_{\ph{j_1/2}0}](6\cdot)\th[^{j_3/4-j_2/3}_{\ph{j_3/4}0}](12\cdot)\sum_{a,b\in\{0,1\}}\th[^{a/2+j_3/4}_{\ph{a/2}0}](4\cdot)\th[^{a/2}_{b/2}]^4(-1)^{b\cdot j_3}\\
&=P_{D_5^{(0)}}\cdot P_{A_3^{(0)}}+2P_{D_5^{(1)}}\cdot P_{A_3^{(1)}}+P_{D_5^{(2)}}\cdot P_{A_3^{(2)}},
\ea
\ee
and for $E_8\supset SU(5)\times SU(5)$ (4 Wilson lines)
\be\label{4wllat}
\ba
\sum_{p\in \Gamma_{E_8}}q^{p^2\over 2}&=\sum_{j_4=0,\cdots,4}\sum_{\begin{subarray}{c}j_1=0,1\\j_2=0,1,2\\j_3=0,\cdots,3\end{subarray}}\th[^{j_1/2}_{\ph{a}0}](2\cdot)\th[^{
j_2/3-j_1/2}_{\ph{j_1/2}0}](6\cdot)\th[^{j_3/4-j_2/3}_{\ph{j_3/4}0}](12\cdot)\th[^{j_4/5-j_3/4}_{\ph{j_4/5}0}](20\cdot)\cdot\\
&\qquad\cdot\sum_{a,B\in\{0,1\}}\th[^{a/2+j_4/5}_{\ph{a/}B/2}](5\cdot)\th[^{a/2}_{B/2}]^3(-1)^{B\cdot j_4}\\
&=P_{A_4^{(0)}}\cdot P_{A_4^{(0)}}+2 P_{A_4^{(1)}}\cdot P_{A_4^{(1)}}+2 P_{A_4^{(2)}}\cdot P_{A_4^{(2)}}.
\ea
\ee
Note, however, that there are many other ways to decompose the lattice under other maximal subgroups. As an example, we can decompose $E_8\rightarrow SO(14)\times SU(2)$ as shown in figure \ref{fig:SO14}:
\begin{figure}[!h]
\centering
\includegraphics[scale=.4]{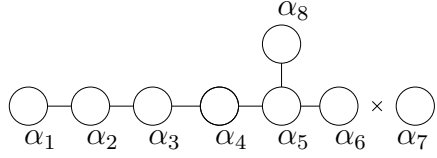}
\caption{The split $E_8\rightarrow SO(14)\times SU(2)$}\label{fig:SO14}
\end{figure}
\be
\sum_{p\in \Gamma_{E_8}}q^{p^2\over 2}=\sum_{j=0,1}\sum_{n_7}q^{(n_7-{j\over 2})^2}\sum_{n_1,\cdots,n_6,n_8}q^{{3\over 4}(2n_6-j)^2+n_8^2+n_5^2\cdots +n_1^2-(2n_6-j)n_5-n_5n_8\cdots-n_7n_8}.
\ee
Denoting the lattice sum $\sum_{p\in\Gamma_{E_8}}q^{p^2\over 2}$ by $f(\tau)$, the splittings \eqref{1wllat}-\eqref{4wllat} labeled by the lower number of Wilson lines \mbox{$k=1,\cdots, 4$} can be cast into the general form
\be\label{latsplit}
f(\tau)=f^k_0\theta^{(8-k)}_{0}+\cdots f^k_k\theta^{(8-k)}_k,
\ee
where
\be
\theta^{(k)}_J:=\sum_{\bs j_1=0,1\\\vdots\\j_{k-1}=0,\cdots k-1\es}\vt[^{j_1\over2}_{\phantom{x}0}](2\cdot)\vt[^{{j_2\over3}-{j_1\over 2}}_{\phantom{{j_2\over3}}0}](6\cdot)\cdots\vt[^{{j_{k-1}\over k}-{j_{k-2}\over {k-1}}}_{\phantom{j_2\over3}0}]((k{\rm -1})\cdot k)\vt[^{{J\over (k+1)}-{j_{k-1}\over k}}_{\phantom{a_2\over3}0}](k\cdot(k{\rm +}1)),
\ee
and 
\be\label{fkgen}
f^k_J=q^{-{kJ^2\over 2(k+1)}}\sum_{\bs p\in\Gamma_{E_8}\\(p,\alpha_1)=\cdots=(p,\alpha_{k-1})=0\\(p,\alpha_k)=J\es}q^{p^2\over 2}.
\ee
For the chains of models in \cite{afiq}, we find the explicit expressions
\be\label{fkev}
f^k_J=\sum_{a,b=0,1}\vt[^{a/2+J/(k+1)}_{\phantom{a/2}b/2}]((k+1)\cdot)\vt[^{a/2}_{b/2}]^{(7-k)}(-1)^{b\cdot J}
\ee
for k even and 
\be\label{fkod}
f^k_J=\sum_{a,b=0,1}\vt[^{a/2+J/(k+1)}_{\phantom{a/2}0}]((k+1)\cdot)\vt[^{a/2}_{b/2}]^{(7-k)}(-1)^{b\cdot J}
\ee
for k odd.\\
We can write down the same decompositions including the shifts due to the orbifold embedding. In the chains of models in \cite{afiq}, the shifts are of the form $\gamma=(\alpha_1+2\alpha_2+\cdots+m\alpha_m)$ and thus deform $p$ to $p+a\gamma=(n_1+a)\alpha_1+(n_2+2a)\alpha_2+\cdots+(n_m+m\cdot a)\alpha_j$. Therefore, $\theta^{(k)}_J$ gets deformed to
\be\label{thetshift}
\ba
&\theta^{(k)}_{J,\gamma}[^a_b](q)=\\
&\sum_{\bs j_1=0,1\\\vdots\\j_{k-1}=0,\cdots k-1\es}\vt[^{{j_1\over2}}_{\phantom{x}0}](2\cdot)\vt[^{{j_2\over3}-{j_1\over 2}}_{\phantom{{j_2\over3}}0}](6\cdot)\cdots\vt[^{{j_m\over(m+1)}-{j_{m-1}\over m}-m\cdot a}_{\phantom{{a_2\over3}}-m(m+1)b}](m\cdot(m+1))\cdots\vt[^{{J\over (k+1)}-{j_{k-1}\over k}}_{\phantom{j_2\over3}0}](k\cdot(k+1)).
\ea
\ee
Similar realizations exist for other types of shifts.
On the part of the lattice denoted by $f^k_J$, it is more convenient to write in an orthogonal basis $\gamma=(\gamma_1,\cdots,\gamma_{7-k},0,\cdots,0)$ and we get for $f^k_J$ with $k$ even
\be\label{fkevshift}
f^k_{J,\gamma}[^a_b]=\sum_{A,B=0,1}\re^{-\pi\ri\sum_i \gamma_i B a}\vt[^{A/2+J/(k+1)}_{\phantom{A/2}B/2}]((k+1)\cdot)\prod_{i=1}^{7-k}\vt[^{A/2+a\gamma_i}_{B/2+b\gamma_i}](-1)^{B\cdot J},
\ee
respectively for $k$ odd,
\be\label{fkodshift}
f^k_{J,\gamma}[^a_b]=\sum_{A,B=0,1}\re^{-\pi\ri\sum_i \gamma_i B a}\vt[^{A/2+J/(k+1)}_{\phantom{A/2}0}]((k+1)\cdot)\prod_{i=1}^{7-k}\vt[^{A/2+a\gamma_i}_{B/2+b\gamma_i}](-1)^{B\cdot J}.
\ee
Cases with more than $7-k$ non-vanishing entries in $\gamma$ have to be considered separately, see section \ref{ssec:moduli}. \\
The lattice splits derived above are the main ingredients for computing the $F^{(g)}$ in models with Wilson lines. Indeed, turning on one Wilson line in the chains of \cite{afiq} corresponds to preserving a $U(1)$ that can be enhanced to an $SU(2)$ while Higgsing an $E_7$, and will therefore be reflected by a split as in \eqref{1wllat}. On the other hand, turning on seven Wilson lines Higgses an $SU(2)$ while preserving a $U(1)^7$ that can be enhanced to $E_7$ and therefore corresponds to the same split with sides exchanged, or equivalently: the same modified Dynkin diagram (Fig. \ref{fig:E81}) with circles replaced by crosses. Similarly, \eqref{2wllat} corresponds to 2, respectively 6 and \eqref{3wllat} to 3, respectively 5 Wilson lines. For 4 Wilson lines, one can choose to Higgs either side of the lattice.

\begin{figure}[!h]
\centering
\includegraphics[scale=.4]{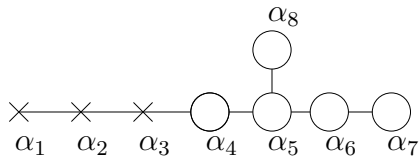}
\caption{$E_8$ with 5 Wilson lines}\label{E85}
\end{figure}
\begin{figure}
\centering
\includegraphics[scale=.4]{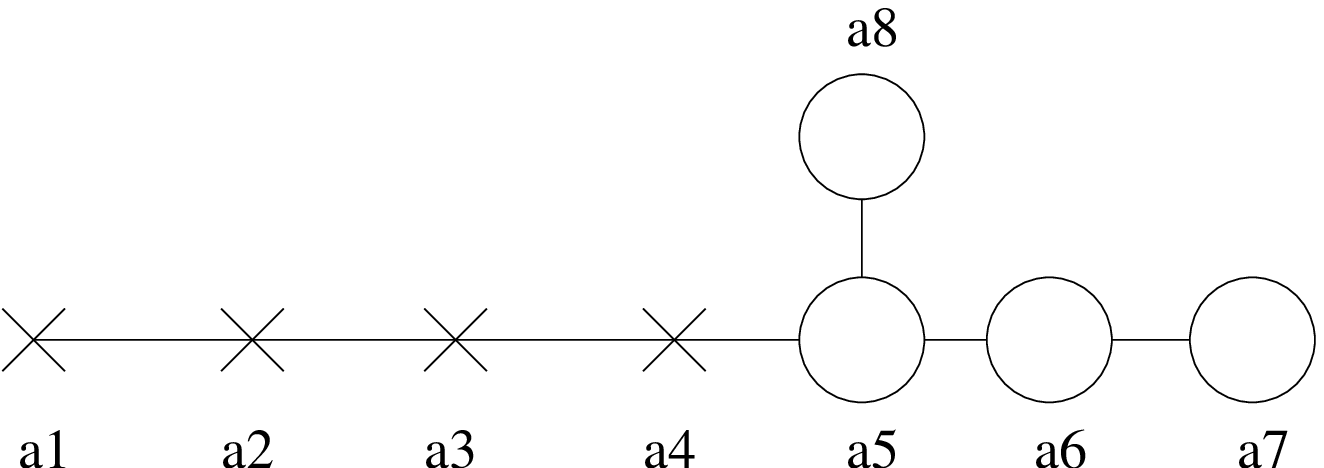}
\caption{$E_8$ with 4 Wilson lines}\label{E84}
\end{figure}
\begin{figure}
\centering
\includegraphics[scale=.4]{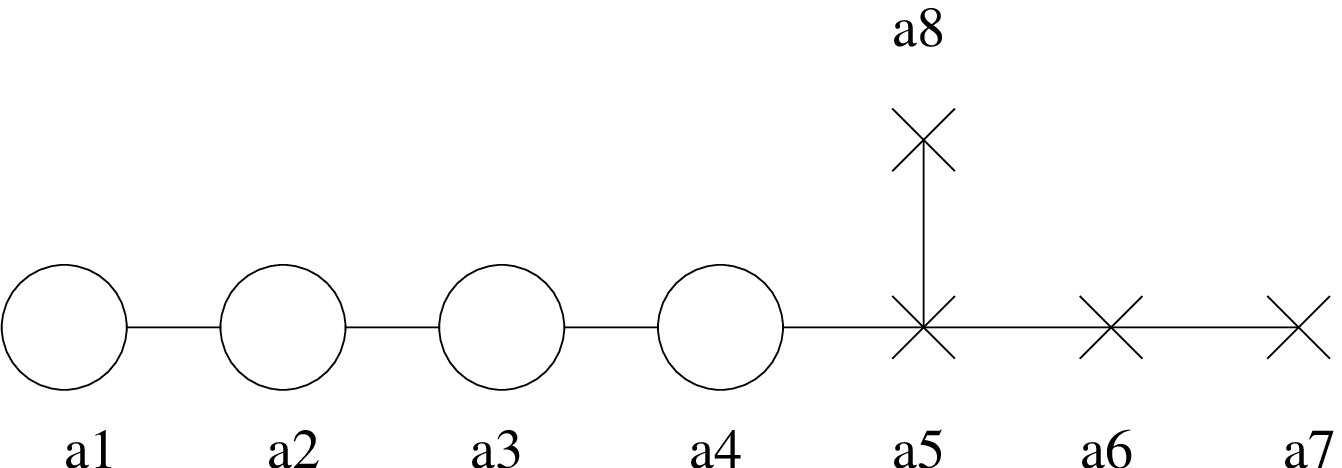}
\caption{$E_8$ with 4 Wilson lines, alternative split}\label{E86}
\end{figure}
\begin{figure}
\centering
\includegraphics[scale=.4]{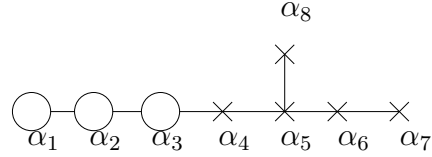}
\caption{$E_8$ with 3 Wilson lines}\label{E83}
\end{figure}
\subsection{Moduli dependence}\label{ssec:moduli}

We can now use the above to decompose the full lattice sum with torus moduli, Wilson moduli, shifts and insertions. Note that when the vector of Wilson line moduli $y$ is \emph{not} orthogonal to the shifts, i.e. $\gamma\cdot y\neq 0$, we turn on Wilson line moduli corresponding to the part of the gauge group only present in the orbifold limit. This results in freezing the vector moduli at that special point of moduli space, and the degeneracy of vacua gets lifted: The couplings corresponding to equivalent embeddings with different N can be different \cite{hm}. \\
We therefore impose here $\gamma\cdot y=0$, restricting the Wilson lines to the part of the lattice orthogonal to the shift. We have to distinguish the cases of less than four Wilson lines from those with four and more. In the latter, $\gamma\cdot y=0$ is automatically fulfilled for the shifts given in table \ref{tab:orb}, as the Wilson lines are active on the right-hand side of the Dynkin diagram while the shifts act on the left. If we turn on less than four Wilson lines, those are active on the left-hand side of the diagram, as explained in section \ref{ssec:split}. This means that we have to choose the shift such that it does not interfere with the Wilson lines, and in such a way that it preserves the part of the diagram where the Wilson lines are active. For the $\mbb{Z}_2,\mbb{Z}_3$ and $\mbb{Z}_4$ embeddings on the first $E_8$ lattice (see table \ref{tab:orb}), it is sufficient to move the shift to the other end of the diagram, redefining $\gamma^1_{\mbb{Z}_2}\rightarrow \gamma'^1_{\mbb{Z}_2}=(0^6,-1,1), \gamma^1_{\mbb{Z}_3}\rightarrow \gamma'^1_{\mbb{Z}_3}=(0^5,-2,1,1),\gamma_{\mbb{Z}_4}\rightarrow \gamma'^1_{\mbb{Z}_4}=(0^4,-3,1,1,1)$. In the case of the $\mbb{Z}_6$ orbifold, this does the trick for one and two Wilson lines, but if we turn on a third one, it is not orthogonal to $\gamma'^1_{\mbb{Z}_6}$ anymore. However, we can choose the equivalent embedding $\gamma'^1=(2,2,2,2,2,0^3)$, orthogonal to $y\in{\rm span}(\alpha_1,\alpha_2,\alpha_3)$. In this case, this is also a valid choice for zero, one and two Wilson lines. The Wilson lines on the second $E_8$, unchanged throughout the sequential Higgs mechanisms, work out similarly. Only the $\mbb{Z}_4$ orbifold is slightly more delicate, as the Wilson lines corresponding to maximal Higgsing on the second $E_8$ preserve an $SO(8)$, and therefore act in the center of the diagram. The combination of theta functions corresponding to the Higgsed lattice can however be determined using \eqref{fkgen}.

For one Wilson line, we thus write
\be
\ba
&\sum_{p\in\Gamma^{18,2}+a\gamma}p_R^{(2g-2)}q^{|p_L|^2\over 2}\bar{q}^{|p_R|^2\over 2}\re^{2\pi\ri b\gamma\cdot p}=\sum_{p\in\Gamma^{18,2}+a\gamma}(p\cdot u(y))^{(2g-2)}q^{p^2\over 2}|q|^{(p\cdot u(y))^2}\re^{2\pi\ri b\gamma\cdot p}\\
&\qquad =\sum_{J=0,1}\sum_{\begin{subarray}{c}A,B\in\{0,1\}\\\alpha,\beta\in\{0,1\}\end{subarray}}\re^{-\pi\ri\sum_i \gamma'_i B a}\left(\prod_{i=3}^8\th[^{A/2+a\gamma'_i}_{B/2+b\gamma'_i}]\right)\th[^{A/2+J/2}_{\ph{A/2}0}](2\cdot)(-1)^{BJ}\\
&\cdot\re^{-\pi\ri a\sum_{i=9}^{16} \gamma_i \beta}\left(\prod_{j=9}^{16}\th[^{\alpha/2+a\gamma_j}_{\beta/2+b\gamma_j}]\right)\cdot\sum_{n_1,n_{\pm},m_{\pm}}(p\cdot u(y))^{2g-2}q^{(n_1-{J\over 2})^2-m_+n_-+n_0m_0}|q|^{(p\cdot u(y))^2}\\
&\qquad=\sum_J f_J^1[^a_b](q)\bar{\Theta}^g_{J,1}(q,y),
\ea
\ee
where $\Theta^g_{J,k}(q,y)$ is defined in \eqref{thetaJ}, and
\be
\ba
f_J^1[^a_b](q)=&\sum_{\begin{subarray}{c}A,B\in\{0,1\}\\\alpha,\beta\in\{0,1\}\end{subarray}}\re^{-\pi\ri a\sum_{i=3}^8 \gamma'_i B}\left(\prod_{i=3}^8\th[^{A/2+a\gamma'_i}_{B/2+b\gamma'_i}]\right)\th[^{A/2+J/2}_{\ph{A/2}0}](2\cdot)(-1)^{BJ}\\
&\cdot\re^{-\pi\ri a\sum_{i=9}^{16} \gamma_i \beta}\left(\prod_{j=9}^{16}\th[^{\alpha/2+a\gamma_j}_{\beta/2+b\gamma_j}]\right).
\ea
\ee
This is nothing else than \eqref{fkodshift} applied to the whole lattice of two $E_8$ and the torus, and including the shifts. Analogously, we get for $k\leq 4$ Wilson lines
\be
\sum_{p\in\Gamma^{18,2}+a\gamma}p_R^{(2g-2)}q^{|p_L|^2\over 2}\bar{q}^{|p_R|^2\over 2}\re^{2\pi\ri b\gamma\cdot p}=\sum_J f_J^k[^a_b](q)\bar{\Theta}^g_{J,k}(q,y),
\ee
where for k=3
\be
\ba
f_J^3[^a_b](q)=&\sum_{\begin{subarray}{c}A,B\in\{0,1\}\\\alpha,\beta\in\{0,1\}\end{subarray}}\re^{-\pi\ri a\sum_{i=5}^8 \gamma'_i B}\left(\prod_{i=5}^8\th[^{A/2+a\gamma'_i}_{B/2+b\gamma'_i}]\right)\th[^{A/2+J/4}_{\ph{A/2}0}](4\cdot)(-1)^{BJ}\\
&\re^{-\pi\ri a\sum_{i=9}^{16} \gamma_i \beta}\left(\prod_{j=9}^{16}\th[^{\alpha/2+a\gamma_j}_{\beta/2+b\gamma_j}]\right),
\ea
\ee
and for $k=2$ or $k=4$ Wilson lines, using \eqref{fkevshift},
\be
\ba
f_J^k[^a_b](q)=&\sum_{\begin{subarray}{c}A,B\in\{0,1\}\\\alpha,\beta\in\{0,1\}\end{subarray}}\re^{-\pi\ri a\sum_{i=k+2}^8 \gamma_i B}\left(\prod_{i=k+2}^8\th[^{A/2+a\gamma'_i}_{B/2+b\gamma'_i}]\right)\th[^{A/2+J/(k+1)}_{B/2}]((k+1)\cdot)(-1)^{BJ}\\
&\re^{-\pi\ri a\sum_{i=9}^{16} \gamma_i \beta}\left(\prod_{j=9}^{16}\th[^{\alpha/2+a\gamma_j}_{\beta/2+b\gamma_j}]\right).
\ea
\ee
When more than four Wilson lines are turned on ($k\geq 4$), we decompose analogously as
\be
\sum_{p\in\Gamma^{18,2}+a\gamma}p_R^{(2g-2)}q^{|p_L|^2\over 2}\bar{q}^{|p_R|^2\over 2}\re^{2\pi\ri b\gamma\cdot p}=\sum_J \theta^k_J[^a_b](q)\bar{\Theta}^g_{J,k}(q,y),
\ee
where $\theta^k_J[^a_b](q)$ is \eqref{thetshift}, supplemented by the contribution from the second $E_8$ lattice.\\
Any other split for any number of Wilson lines fulfilling the constraint $\gamma\cdot y=0$ can be realized similarly. In the above, we have assumed that the second $E_8$ lattice is Higgsed completely, without any Wilson lines. If this is not the case, as for example for the $\mbb{Z}_2,\mbb{Z}_3$ and $\mbb{Z}_4$ models in \cite{afiq}, the second lattice also has to be split according to the above prescription. 

Note that these splits describe a ``generalized hatting procedure'' analogous to the 1-Wilson line case analyzed in \cite{ccl} for generalized Jacobi forms. In the 1 Wilson line $STUV$ model, the relevant forms are standard Jacobi forms 
\be
f(\tau,V)=\sum_{\bs n\geq 0\\l\in\mbb{Z}\es}c(4n-l^2)q^n r^l
\ee
with $q=\re^{2\pi\ri\tau},r=\re^{2\pi V}$, admitting a decomposition
\be
f(\tau,V)=f_{ev}(\tau)\theta_{ev}(\tau,V)+f_{odd}(\tau)\theta_{odd}(\tau,V),
\ee
where $\theta_{ev}=\theta_3(2\tau,2V)$, $\theta_{odd}=\theta_2(2\tau,2V)$.
The effect of turning on a Wilson line can be described by replacing $f(\tau,V)$ by its hatted counterpart \cite{ccl}
\be
\hat{f}(\tau,V)=f_{ev}(\tau)+f_{odd}(\tau)
\ee
In the generic, k Wilson line case considered here, we decompose the lattice sum as in \eqref{latsplit}.

 When $k\leq 4$, the ``generalized hatting'' due to the Wilson lines is
\be
\hat{f}[^a_b](\tau,V_1,\cdots V_k)=f^k_0[^a_b](\tau)+\cdots f^k_k[^a_b](\tau),
\ee
where $f^k_J$ and $f^k_{k+1-J}$ are equivalent.
When $k\geq 4$, we have to keep the other part of the split lattice. This yields the ``complementary hatting''
\be
\breve{f}(\tau,V_1,\cdots V_n)=\theta^{8-k}_0[^a_b](\tau)+\cdots \theta^{8-k}_k[^a_b](\tau),
\ee
with $\theta^{8-k}_J=\theta^{8-k}_{k+1-J}$.

\subsection{Computation of $F^{(g)}$}
 In the following, we will denote the number of Wilson lines by k and write the split lattice sum as
\be
\sum_J \Phi_J^k[^a_b](q)\bar{\Theta}^g_{k,J}(q),
\ee
where $\Phi_J^k[^a_b](q)$ is the function appearing in \eqref{int3} and stands for $f^k_J[^a_b]$ or $\theta_J^k[^a_b](q)$, whichever is applicable.
We expand the modular function in the integrand of \eqref{int3} as 
\be\label{exp}
\mathcal{P}_{2g}(q)\CF^k_J(q):=\mathcal{P}_{2g}(q)\sum_{a,b}{c(a,b)\re^{2\pi\ri ab(2-\gamma^2)}\over \eta^{18}\vt[^{1+a}_{1+b}]\vt[^{1-a}_{1-b}]}\Phi_J^k[^a_b](q)=\sum_{n\in\mbb{Q}_J}c_{g,J}^k(n)q^{n},
\ee
where $\mbb{Q}_J$ denotes the subset of $\mbb{Q}$ containing the powers of q appearing in the conjugacy class $J$. Since different conjugacy classes correspond to different rational powers of $q$, we can sum over $J$ without loss of information and write
\be
\sum_{n\in\mbb{Q}} c_g^k(n)q^n=\sum_J\sum_{n\in\mbb{Q}_J}c_{g,J}^k(n)q^{n}.
\ee
We can now evaluate the integral \eqref{int} using Borcherds' technique of lattice reduction \cite{borcherds} reviewed in appendix \ref{ap:latred}. We choose the reduction vector to lie in the torus part of the lattice, the result is therefore only valid in the chamber of the $T,U$ torus moduli space where the projected reduction vector $z_+$ is small. The result looks very similar to what was obtained in \cite{mm2} for the STU-model and can be simplified to read \footnote{see the appendix of \cite{gkmw} for details of the simplification} $F^{(g)}=F^{(g)}_{\rm deg}+F^{(g)}_{\rm nondeg}$ where
\be\label{fdeg}
F^{(g)}_{\rm deg}={(y_2,y_2)8\pi^3\over T_2}\delta_{g,1}+{1\over 2(2T_2)^{2g-3}}\sum_{\lambda\in\Gamma^{k,0}}\sum_{l=0}^g\Li_{2l-2g+4}(q^{\RE(\bar{\lambda}\cdot\bar{y})})c_{g-l}^k({\lambda^2\over 2}){1\over \pi^{2l+3}}(-{T_2^2\over 2y_2^2})^l
\ee
\be\label{fnondeg}
\ba
&F^{(g)}_{\rm nondeg}=\\
&\sum_{l=0}^{g-1}\sum_{C=0}^{\begin{subarray}{c}{\rm min}\\(l,2g-3-l)\end{subarray}}\sum_{r\in\Gamma^{k+1,1}}\binom{2g-l-3}{C}{1\over (l-C)!2^C}{(-\RE(r\cdot y))^{l-C}\over (y_2,y_2)^l}c_{g-l}^k({r^2\over 2})\Li_{3-2g+l+C}(\re^{-r\cdot y})\\
&\qquad+{c_1^k(0)\over 2^g(g-1)(y_2,y_2)^{g-1}}+\sum_{l=0}^{g-2}{c_{g-l}^k(0)\over l!(2(y_2,y_2))^l}\zeta(3+2(l-g)){(2g-3-l)!\over (2g-3-2l)!}
\ea
\ee
This can also be compared to the expressions obtained in \cite{hm} for genus one. The lattice sum in \eqref{fnondeg} is over the so-called reduced lattice $\Gamma^{k+1,1}$. This is a sublattice of the original lattice $\Gamma^{k+2,2}$, parametrized by $(n_0,m_0,b_i)$.\\
A highly nontrivial check of the computation is provided by the Euler characteristics of the corresponding Calabi-Yau manifolds, respectively the difference $n_h-n_v$ on the heterotic side. Heterotic-type II duality implies \cite{mm2} that it should be given by the normalized $q^0$ coefficient of $\CF_J^k$, namely
\be
2(n_h-n_v)=\chi(X)=2{c_0^k(0)\over c_0^k(-1)}.
\ee
One indeed finds precisely the chains of Euler characteristics given in \cite{afiq}, see table \ref{tab:Euler}. The corresponding K3-fibrations are listed in table \ref{tab:dualCY}.

\begin{table}[!h]
\begin{center}\begin{tabular}{|c| c c c c c c c c|}
\hline
$\mbb{Z}_2$ & 92 & 132 & 168 & 200 & 304 & 412 & 612 & 960\\
$\mbb{Z}_3$ &    & 120 & 144 & 164 & 232 & 312 & 420 & 624 \\
$\mbb{Z}_4$ &    &    &    &     & 224 & 288 & 372 & 528\\
$\mbb{Z}_6$ &    &    &    & 220 & 264 & 312 & 372 &  480\\
\hline
\end{tabular}\end{center}
\caption{Euler characteristics $\chi$ for the models in \cite{afiq}}\label{tab:Euler}
\end{table}

\section{Heterotic-type II duality and instanton counting}\label{sec:dual}
\subsection{Moduli map}
In this section, we will determine geometric quantities on the dual Calabi-Yau manifolds on the type II side using the heterotic expressions obtained above.\\
The heterotic dilaton $S$ gets mapped to the K\"ahler modulus $t_2$, therefore heterotic weak coupling regime corresponds to $t_2\rightarrow\infty$. This restricts the instanton numbers accessible to our computation to those classes where the corresponding coefficient $l_2$ vanishes. The mapping of the remaining heterotic moduli from the Torus and the Wilson lines $(T,U,V_1,\cdots V_k)$ to the K\"ahler moduli $(t_1,\cdots t_{k+3})$ on the type II side can be determined for models with small number of K\"ahler moduli comparing the classical pieces of the prepotential \cite{ccl}. In order to compare with the instanton numbers in \cite{bkkm}, we extend the map of \cite{ccl} to two Wilson lines as follows:  
\be
\ba
T&\rightarrow t_1+2t_4+3t_5\\
U&\rightarrow t_1+t_3+2t_4+3t_5\\
V_1&\rightarrow t_4\\
V_2&\rightarrow t_5
\ea
\ee
implying that the numbers $(n_{0},m_{0},b_i)$ in \eqref{param} map to the numbers $l_i$ on the type II side as
\be\label{map}
\begin{array}{lll}
l_1&=n_0+m_0\qquad\qquad &l_4=2(n_0+m_0)+b_1\\
l_2&=0\qquad &l_5=3(n_0+m_0)+b_2\\
l_3&=n_0.
\end{array}
\ee
For higher numbers of Wilson lines, we cannot conclusively determine the map due to lack of information on the type II side, but it is clear that such a map exists and that it is linear.
 In order to extract genus g instanton numbers from the expansion \eqref{exp}, we have to specify the norm $(p,p)$. Redefining the indices in \eqref{1wllat}-\eqref{4wllat} as 
\be
\ba
&(n_1-{a\over 2})^2 &\rightarrow &\ {b_1^2\over 4}\\
&(n_1-{a\over 2})^2+3(n_2+{a\over 2}-{b\over 3})^2 &\rightarrow &\  {b_1^2\over 4}+3({b_1\over 2}-{b_2\over 3})^2=b_1^2-b_1b_2+{b_2^2\over 3}\\
&(n_1-{a\over 2})^2+3(n_2+{a\over 2}-{b\over 3})^2+6(n_2+{b\over 3}-{c\over 4})^2 &\rightarrow &\ {b_1^2\over 4}+3({b_1\over 2}-{b_2\over 3})^2+6({b_2\over 3}-{b_3\over 4})^2\\
&&=&\ b_1^2+b_2^2-b_1b_2-b_2b_3+{3b_3^2\over 8},\\
&&\vdots &
\ea
\ee
we find the norms given in table \ref{tab:norm}. We thus have for the instanton numbers
\be
\ba
&c^g_k(n_0,m_0,b_1,\cdots b_k)=c^g_k(n_0m_0-b_1^2-\cdots-b_{k-1}^2+b_1b_2\cdots b_{k-1}b_k-{k b_k^2\over 2(k+1)}),\qquad k\leq 4\\
&c^g_k(n_0,m_0,b_{9-k},\cdots b_8)=c^g_k(n_0m_0-{(10-k)b_{9-k}^2\over 2(9-k)}-b_{10-k}^2-\cdots-b_8^2+b_{9-k}b_{10-k}+\cdots b_5b_8,\\&\hspace{14.8cm}k\geq 4,\\
\ea
\ee
confirming the conjecture made in \cite{ccl}. Note that the last $b_p$ determines the conjugacy class.

\begin{table}[!h]
\begin{center}\begin{tabular}{|l | l|}
\hline
k & $p_{\rm het}^2$\\
\hline
$0$&$n_0m_0$\\
$1$&$n_0m_0-{b_1^2\over 4}$\\
$2$&$n_0m_0-b_1^2+b_1b_2-{b_2^2\over 3}$\\
$3$&$n_0m_0-b_1^2-b_2^2+b_1b_2+b_2b_3-{3b_3^2\over 8}$\\
$4$&$n_0m_0-b_1^2-b_2^2-b_3^2+b_1b_2+b_2b_3+b_3b_4-{2b_4^2\over 5}$\\
$5$&$n_0m_0-{5b_4^2\over 8}-b_5^2-b_6^2-b_7^2-b_8^2+b_4b_5+b_5b_6+b_5b_8+b_6b_7+b_7b_8$\\
$6$&$n_0m_0-{2b_3^2\over 3}-b_4^2-b_5^2-b_6^2-b_7^2-b_8^2+b_3b_4+b_4b_5+b_5b_6+b_5b_8+b_6b_7+b_7b_8$\\
$7$&$n_0m_0-{3b_2^2\over 4}-b_3^2-b_4^2-b_5^2-b_6^2-b_7^2-b_8^2+b_2b_3+b_3b_4+b_4b_5+b_5b_6+b_5b_8+b_6b_7+b_7b_8$\\
\hline
\end{tabular}
\end{center}
\caption{The norm $(p_{\rm het},p_{\rm het})_k$ for $k=(0,1,\cdots 7)$ Wilson lines}\label{tab:norm}
\end{table}
\subsection{Extracting geometric information}
The topological couplings $F^{(g)}$ are the free energies of the A-model topological string. They have a geometric interpretation as a sum over instanton sectors,
\be
F^{(g)}(t)=\sum_\beta N_{g,\beta}Q^\beta,
\ee
where $Q_i=\re^{-t_i}$, $\beta=\{n_i\}$ in a basis of $H_2(X)$ denotes a homology class, $Q^\beta:=\re^{-t_in_i}$, and $N_{g,\beta}$ are the Gromov-Witten invariants, in general \emph{rational} numbers. With the work of Gopakumar and Vafa \cite{gv}, a hidden integrality structure of the $N_{g,\beta}$ has been uncovered. The generating functional of the $F^{(g)}$,
\be
F(t,g_s)=\sum_{g=0}^\infty F^{(g)}(t)g_s^{2g-2},
\ee
can be written as a generalized index counting BPS states in the corresponding type \mbox{IIA} theory:
\be
F(t,g_s)=\sum_{g=0}\sum_{\beta}\sum_{d=1}^{\infty}n^g_\beta {1\over d}\left(2\sin{dg_s\over 2}\right)^{2g-2}Q^{d\beta},
\ee
where the numbers $n^g_\beta$ are now \emph{integers} called Gopakumar-Vafa invariants. Since the homology classes $\beta$ are labeled by lattice vectors $p$, we write the Gopakumar-Vafa invariants for models with k Wilson lines as $n^k_g(p)\equiv n^k_g({p^2\over 2})$. We also write, in terms of the instanton degrees on the type II side, $n^k_g(l_1,\cdots,l_{k+3})$.\\
From the structure of the $F^{(g)}$, one can deduce that the coefficients $c^k_{g}({p^2\over 2})$ appearing in \eqref{fdeg},\eqref{fnondeg} are related to the Gopakumar-Vafa invariants through
\be
\sum_{g\geq 0}n^k_g(p)\left(2\sin{\lambda\over 2}\right)^{2g-2}=\sum_{g\geq 0}c^k_g({p^2\over 2})\lambda^{2g-2}.
\ee
The Gopakumar-Vafa invariants can be obtained efficiently using the formula \cite{km}
\be
\sum_{p\in {\rm Pic}(K3)}\sum_{g=0}^\infty n^k_g(p)z^gq^{p^2\over 2}=\sum_J \CF^k_J(q)\xi^2(z,q),
\ee
where $\CF^k_J(q)$ is defined in \eqref{exp}, and
\be
\xi(z,q)=\prod_{n=1}^\infty{(1-q^n)^2\over (1-q^n)^2+zq^n}.
\ee
\subsection{Gopakumar-Vafa invariants}
Table \ref{tab:stu}- table \ref{tab:stuvw} show conjectural GV invariants $n^k_g$ for the K3 fibrations dual to the $STU$-, the $STUV$-, and the $STUV_1V_2$-model. Similar tables for the other models considered in this work can be found in appendix \ref{ap:data}, along with a list of the dual pairs of \cite{afiq}.
\begin{table}[!h]
\begin{center}\begin{tabular}{|r|r|r|r|r|r|r|r|r|}
\hline
$g$& ${p^2\over 2}=-1$&$0$&$1$&$2$&$3$&$4$&$5$\\
\hline
0&-2&480&282888&17058560&477516780&8606976768&115311621680\\
\hline
1&0&4&-948 &-568640 &-35818260 & -1059654720& -20219488840\\
\hline
2&0&0&-6&1408&856254&55723296&1718262980\\
\hline
3&0&0&0&8&-1860 &-1145712 &-76777780\\
\hline
4&0&0&0&0&-10&2304&1436990\\
\hline
\end{tabular}
\caption{$n^k_g({p^2\over 2})$ for $\mbb{Z}_6$, 0 Wilson lines (STU), dual to $X^{1,1,2,8,12}$}\label{tab:stu}
\vspace{.5cm}
\begin{tabular}{|r|r|r|r|r|r|r|r|r|r|r|}
\hline
$g$& ${p^2\over 2}=-1$&$-{1\over 4}$&$0$&${3\over 4}$&$1$&${7\over 4}$&$2$&${11\over 4}$&$3$\\
\hline
0&-2&56& 372& 53952 &174240 &3737736 & 9234496 & 110601280& 237737328\\
\hline
1&0&0&4&-112&-732&-108240&-350696& -7799632 & -19517380\\
\hline
2&0&0&0&0&-6&168& 1084& 162752&528582\\
\hline
3&0&0&0&0&0&0&8& -224&-1428\\
\hline
\end{tabular}
\caption{$\mbb{Z}_6$,1 Wilson line (STUV), dual to $X^{1,1,2,6,10}$}\label{tab:stuv}
\end{center}
\end{table}
\begin{table}
\begin{center}
\begin{tabular}{|r|r|r|r|r|r|r|r|r|r|r|}
\hline
$g$& ${p^2\over 2}=-1$&$-{1\over 3}$&$0$&${2\over 3}$&$1$&${5\over 3}$&$2$&${8\over 3}$&$3$\\
\hline
0&-2& 30& 312 &26664 &120852 & 1747986 & 5685200 & 49588776 & 135063180\\
\hline
1&0&0&4&-60 &-612 &-53508&-243560&-3656196&-12097980\\
\hline
2&0&0&0&0&-6& 90& 904& 80472& 367458\\
\hline
3&0&0&0&0&0&0&8&-120&-1188\\
\hline
4&0&0&0&0&0&0&0&0&-10\\
\hline
\end{tabular}
\caption{$\mbb{Z}_6$, 2 Wilson lines (STU$V_1V_2$), dual to $X^{1,1,2,6,8}$}\label{tab:stuvw}
\end{center}
\end{table}

For comparison with \cite{bkkm}, we give the genus 0 instanton numbers in notation\\ 
\mbox{$[l_1\cdots l_{k+3}]=n^k_0(l_1,\cdots l_{k+3})$} for models with one and two Wilson lines in table \ref{tab:InstZ61}, \ref{tab:InstZ62}. We find indeed perfect agreement with \cite{bkkm}.
\begin{table}[!h]
\begin{center}\begin{tabular}{|l | l|l|l|}
\hline
$[0001]\quad 56$&$[1001]\quad 56$&
$[1003]\quad 56$&$[3014]\quad $174240\\
$[0002]\quad $-2&$[1002]\quad 372$&
$[1000]\quad $-2&$[1011]\quad 56$\\
$[1004]\quad $-2&$[2012]\quad 372$&
$[0003]\quad 0$&$[2013]\quad 53952$\\
\hline
\end{tabular}\end{center}
\caption{Numbers of rational curves of degree $[l_1,0,l_2,l_3,l_4]$ on $X^{1,1,2,6,10}$ (dual to $\mbb{Z}_6$,1 WL)}\label{tab:InstZ61}
\end{table}
\begin{table}[!h]
\begin{center}\begin{tabular}{|l | l|l|l|}
\hline
$[00001]\quad 30$&$[10011]\quad  $30&$[00002]\quad 0$&$[10023]\quad $312\\
$[00010]\quad $-2&$[10022]\quad$30&$[00012]\quad $30&$[10010]\quad $-2\\
$[00023]\quad $-2&$[20101]\quad $26664&$[00011]\quad $30&$ [20169]\quad $312\\
$[00101]\quad$0&$[30141]\quad $0&$[00013]\quad $-2&$[30144]\quad$30\\
$[30145]\quad$26664&$[30146]\quad $120852&$[30147]\quad $26664&$[30148]\quad$30\\

\hline
\end{tabular}\end{center}
\caption{Numbers of rational curves of degree $[l_1,0,l_3,l_4,l_5]$ on $X^{1,1,2,6,8}$ (dual to $\mbb{Z}_6$, 2 WL)}\label{tab:InstZ62}
\end{table}

Another nontrivial check is provided by the requirement of consistent truncation: in \cite{bkkm}, the authors deduce that the following relations have to hold between instanton numbers with 3,4,and 5 moduli
\be
n^0_0(l_1,l_2,l_3)=\sum_x n^1_0(l_1,l_2,l_3,x)\hspace{2cm} n^1_0(l_1,l_2,l_3,l_4)=\sum_x n^2_0(l_1,l_2,l_3,l_4,x).
\ee
Our numbers indeed fulfill this constraint, as for example
\be
n^2_0(0,0,0,1,0)+\cdots +n^2_0(0,0,0,1,3)=-2+30+30-2=56=n^1_0(0,0,0,1),
\ee
\be
n^1_0(0,0,0,0)+\cdots + n^1_0(0,0,0,4)=-2+56+372+56-2=480=n^0_0(0,0,0),
\ee
and
\be
n^2_0(3,0,1,4,0)+\cdots+n^2_0(3,0,1,4,8)=174240=n^1_0(3,0,1,4).
\ee
This relation should also hold at higher genus and for higher numbers of K\"ahler moduli \cite{kkrs}, namely we expect
\be
n^k_g(l_1,l_2,\cdots l_{k+3})=\sum_x n^{k+1}_g(l_1,l_2,\cdots l_{k+3},x).
\ee
Indeed, we have for example for truncation from 2 to 1 Wilson line (tables \ref{tab:stuv}, \ref{tab:stuvw}) $4-60-60+4=-112$, $-6+90+90-6=168$, and $90+904+90=1084$. All instanton numbers produced, including those in tables \ref{tab:first}-\ref{tab:last}, fulfill the truncation identities
\be
\ba
n^0_g(1)&=2\left(n^1_g(0)+n^1_g({3\over 4})\right)+n^1_g(1)\qquad &n^0_g(2)&=2\left(n^1_g(-{1\over 4})+n^1_g(1)+n^1_g({7\over 4})\right)+n^1_g(2)\\
n^1_g(1)&=2\left(n^2_g(-{1\over 3})+n^2_g({2\over 3})\right)+n^2_g(1)\qquad &n^1_g(2)&=2\left(n^2_g(-1)+n^2_g({2\over 3})+n^2_g({5\over 3})\right)+n^2_g(2)\\
n^2_g(1)&=2\left(n^3_g(-{1\over 2})+n^3_g({5\over 8})\right)+n^3_g(1)\qquad &n^2_g(2)&=2\left(n^3_g({1\over 2})+n^3_g({13\over 8})\right)+n^3_g(2)\\
n^2_g({2\over 3})&=n^3_g(-{3\over 8})+n^3_g(0)+n^3_g({1\over 2})+n^3_g({5\over 8}) &n^3_g(0)&=n^4_g(-{2\over 5})+n^4_g(0).
\ea
\ee
Note that these identities hold --as far as we can verify-- at general genus and independently of the specific chain, as expected. Again, this provides a non-trivial check of our results.
\section{Conclusion}\label{sec:concl}
We have shown how to compute higher derivative couplings for general symmetric $\mbb{Z}_N$, $\CN=2$ orbifold compactifications of the heterotic string with any number of Wilson lines. In particular, this provides conjectural instanton numbers for any of the models in the chains of heterotic-type II duals of \cite{afiq}.\\
Unfortunately, our results can so far only be checked for up to two Wilson lines, since for higher numbers of vector multiplets the type II computation becomes very involved. They do however fulfill nontrivial constraints coming from the geometric transitions on the type II side \cite{bkkm}.\\
Furthermore, a rigorous mathematical framework for computing Gromov-Witten invariants along the fiber of certain K3-fibrations has been established in \cite{mp1,mp3}. With these techniques, one might be able to prove some of our physical predictions for Calabi-Yau manifolds of this type.\\
The computation is rather general and might be applicable to other models, e.g. to asymmetric orbifolds.\\
\section*{Acknowledgments}
I would like to thank M.~Mari\~no for suggesting the topic, many helpful comments and discussions, and a critical reading of the manuscript. I also thank E.~Scheidegger for discussions, J.~David for comments on related topics, and especially S.~Stieberger for very valuable remarks. The major part of this work was carried out under the Marie Curie EST program.
\vspace{2cm}
\appendix

\noindent {\bf \Large Appendices}

\section{Theta functions}\label{ap:thet}
\vskip .5cm
{\bf Properties}\vskip .4cm
In our conventions, the theta functions are defined as follows:

\be
\th[^a_b](v|\t)=\sum_{n\in Z}q^{{1\over 2}\left(n-a\right)^2}
e^{2\pi i\left(v-b\right)\left(n-a\right)}\label{t1}
\ee
where $a,b$ are rational numbers and $q=e^{2\pi i\t}$.

\vskip .5cm
They show the following periodicity properties:

\be
\th[^{a+1}_{\phantom{+}b}](v|\t)=\th[^a_b](v|\t)\;\;\;,\;\;\;
\th[^{\phantom{+}a}_{b+1}](v|\t)=e^{2i\pi a}\th[^a_b](v|\t)
\,,\label{t2}\ee
\be
\th[^{-a}_{-b}](v|\t)=\th[^{a}_{b}](-v|\t)\;\;\;,\;\;\;
\th[^a_b](-v|\t)=
e^{4i\pi ab}\th[^a_b](v|\t) ~~~(a,b\in Z)
\,.\label{t3}\ee

We will use a modified Jacobi/Erderlyi  notation where $\th_1=\th[^{1/2}_{1/2}]$,
$\th_2=
\th[^{1/2}_0]$, $\th_3=\th[^0_0]$, $\th_4=\th[^0_{1/2}]$.

\vskip .5cm
Under modular transformations, the theta functions transform according to

\be
\th[^a_b](v|\t+1)=e^{-i\pi a(a-1)}~\th[^{\phantom{a+}a}_{a+b-1/2}](v|\t)
\,,\label{t4}\ee
\be
\th[^a_b]\left({v\over \t}|-{1\over \t}\right)=\sqrt{-i\t}~
e^{2i\pi ab+i\pi{v^2\over \t}}~\th[^{\phantom{*}b}_{-a}](v|\t)
\,.\label{t5}\ee

The Dedekind $\eta$-function of weight ${1\over 2}$ is related to the v-derivative of $\th_1$:
\be
\eta(\t)=q^{1\over 24}\prod_{n=1}^{\infty}(1-q^n),\label{t10}\ee
\be
{\partial\over \partial v}\th_1(v)|_{v=0}\equiv \th_1'=2\pi\eta^3(\t).
\label{t11}\ee
We can always set the variable $v$ to zero by changing the shifts $(a,b)$ appropriately:

\be
\th[^a_b]\left(v+\epsilon_1\t+\epsilon_2|\t\right)=
e^{-i\pi\t\epsilon_1^2-i\pi\epsilon_1(2v-b)-
2i\pi\epsilon_1\epsilon_2}
{}~\th[^{a-\epsilon_1}_{b-\epsilon_2}](v|\t)
\,.\label{t12}\ee
In our conventions, we will systematically use shifts rather than the variable $v$.
\vskip .5cm
We also note the following identities

\be
\th_2(0|\t)\th_3(0|\t)\th_4(0|\t)=2~\eta^3
\,,\label{t13}\ee

\be
\th_2^4(v|\t)-\th_1^4(v|\t)=\th_3^4(v|\t)-\th_4^4(v|\t)
\,,\label{t14}\ee

We have the following identities for the derivatives of $\th$-functions
\be\partial_\t({\th_2\over \eta})={i\pi\over 12 \eta}
\left(\th_3^4+\th_4^4\right)
\ee
\be
\partial_\t({\th_3\over \eta})={i\pi\over 12 \eta}
\left(\th_2^4-\th_4^4\right)
\ee
\be
\partial_\t({\th_4\over \eta})={i\pi\over 12 \eta}
\left(-\th_2^4-\th_3^4\right)
\ee
Note that the above is valid for all rational values of a,b,h,g. The case $h,g\in\{0,1/2\}$can be seen as a special case, relevant for $\mbb{Z}_2$-orbifolds, while $h,g\in\{0,1/n,\cdots (n-1)/n\}$ arise in the $\mbb{Z}_n$-case (see, e.g., \cite{mumford} or \cite{a-gmv}). \\
We also use the short-hand notation
\be
\vt[^a_b](\tau):=\vt[^a_b](0|\tau)
\ee
as well as
\be
\vt[^a_b](m\cdot):=\vt[^a_b](0|m\tau)
\ee
\vskip .8cm
{\bf Eisenstein series}\vskip .4cm
The Eisenstein series $E_{2n}$ are defined as
\be
E_{2n}=1-{4n\over B_{2n}}\sum_{k\geq 1} {k^{2n-1}q^k\over 1-q^k}.
\ee
$E_{2n}$ with $n>1$ are holomorphic modular forms of weight $2n$. The Eisenstein series $E_2$ is often called quasi modular since under modular transformations, it transforms with a shift
\be
E_2(-{1\over \tau})=\tau^2 \left(E_2(\tau)+{6\over \pi\ri\tau}\right).
\ee
Adding a term that compensates this shift yields the modular, but only ``almost holomorphic'' form of weight two $\widehat{E}_2$
\be
\widehat{E}_2=E_2-{3\over \pi\tau_2}.
\ee
The ring of almost holomorphic modular forms is generated by $\widehat{E}_2$ and the next two Eisenstein series
\be
\ba
E_4&=1+240\sum_{k\geq 1}{k^3q^k\over 1-q^k}={1\over 2}\sum_{a,b}\vt[^a_b]^8\\
E_6&=1-504\sum_{k\geq 1}{k^5q^k\over 1-q^k}.
\ea
\ee
\vskip .8cm
{\bf Lie algebra lattice sums}\vskip .4cm
Any shifted lattice sum over $E_8$ can be written in terms of theta functions as
\be\label{latsum}
\sum_{p\in\Gamma_{E_8}+a\gamma}q^{p^2\over 2}\re^{2\pi\ri bp\cdot\gamma}=\sum_{\alpha,\beta}\prod_{i=1}^8\vt[^{\alpha+a\gamma_i}_{\beta+b\gamma_i}]\re^{-\pi\ri\sum_i \gamma_i \beta a}
\ee
In particular,
\be
E_4={1\over 2}\sum_{p\in\Gamma_{E_8}}q^{p^2\over 2}
\ee
and $E_6$ is related to the $E_8$ lattice shifted by any modular invariant embedding $\gamma$
\be
E_6=\sum_{(a,b)\neq (0,0)}{c(a,b)\over 2\vt[^{{1\over 2}+a}_{{1\over 2}+b}]\vt[^{{1\over 2}-a}_{{1\over 2}-b}]}\sum_{p\in\Gamma_{E_8}+a\gamma}q^{p^2\over 2}\re^{2\pi\ri bp\cdot\gamma},
\ee
with $c(a,b)$ as defined in section \ref{sec:coupl}.\\
An obvious generalization of \eqref{latsum} is the modified Siegel-Narain Theta function over a general shifted lattice $\Gamma$ of signature $(b^+,b^-)$ with an insertion of $(p_R)^{2g-2}$
\be
\Theta_\Gamma^g(\tau,\gamma,a,b)=\sum_{p\in\Gamma+a\gamma}(p_R)^{2g-2}q^{|p_L|^2\over 2}\bar{q}^{|p_R|^2\over 2}\re^{2\pi\ri b \gamma\cdot p}.
\ee
We also use the notation
\be
\Theta_{\Gamma}(\tau,\gamma_1,\gamma_2;P,\phi)=\sum_{p\in\Gamma+\gamma_1}{\rm exp}(-{\Delta\over 8\pi\tau_2})\phi(P(p))q^{|p_L|^2\over 2}\bar{q}^{|p_R|^2\over 2}\re^{2\pi\ri \gamma_2\cdot p},
\ee
where $\gamma_1,\gamma_2$ are shifts, P is an isometry from $\Gamma\times\mbb{R}$ to $\mbb{R}^{b^+,b^-}$, $\phi$ is a polynomial on $\mbb{R}^{b^+,b^-}$ of degree $m^+$ in the first $b^+$ variables and of degree $m^-$ in the others, and $\Delta$ is the Euclidean Laplacian on $\mbb{R}^{b^+,b^-}$. The isometry P defines projections on $\mbb{R}^+,\mbb{R}^-$ written as $P_+(p)=p_R,\quad P_-(p)=p_L$. We will here only consider cases where the shifts are proportional, $\gamma_1=a\gamma\sim\gamma_2=b\gamma$.
\section{Lattice reduction}\label{ap:latred}
In \cite{borcherds}, Borcherds developed the technique of lattice reduction to compute integrals of the form
\be\label{intlat}
\Phi_\Gamma=\int_\CF {d^2\tau\over \tau_2^2} F_M(\tau)\Theta_M(\tau,\gamma_1,\gamma_2;P,\phi),
\ee
where M is a lattice of signature $(b^+,b^-)$, $\Theta_M(\tau,\gamma_1,\gamma_2;P,\phi)$ is the generalized Siegel theta function with projection $P$ and polynomial insertion $\phi$ as defined in appendix \ref{ap:thet} and $F_M$ is a (quasi) modular form of weight $(-{b^-\over 2}-m^-,-{b^+\over 2}-m^+)$ that can be constructed from a (quasi) modular form $F$ with weights $({b^+\over 2}+m^+-{b^-\over 2}-m^-,0)$ as $F_M=\tau_2^{{b^+\over 2}+m^+}F$.
The integral \eqref{intlat} can be decomposed into a sum over a reduced lattice $K$ of signature $(b^+-1,b^--1)$ and a new integral $\Phi_K$ involving $K$ instead of $M$ (\cite{borcherds}, Theorem 7.1). Iterating this procedure, on arrives at an integral $\Phi_{K_f}$ with a lattice $K_f$ of signature $(b^+-b^-,0)$ respectively $(0,b^--b^+)$ that can in principle be solved using standard methods. \\
The reduction steps proceed as follows. Choose two vectors $z$, $z'$ in $M$ with z primitive and $(z,z)=0$, $(z,z')=1$. The reduced lattice is then defined as $K={M\ \cap\ z^{\perp}\over \mbb{Z}_z}$. We also define reduced projections $\tilde{P}$ in a natural way: 
\be
\tilde{P}_{\pm}(\lambda)=P_{\pm}(\lambda)-{(P_{\pm}(\lambda),z_{\pm})\over z_{\pm}^2}z_\pm.
\ee
We can then expand the polynomial $\phi$ in terms of $(\lambda,z_\pm)$ as
\be
\phi(P(\lambda))=\sum_{h^+,h^-}=(\lambda,z_+)^{h^+}(\lambda,z_-)^{h^-}\phi_{h^+,h^-}(\tilde{P}(\lambda)).
\ee
The statement of Borcherds' theorem is then that with these conventions, $z_+^2$ \emph{sufficiently small} and $\tilde{P}_+(\lambda_K)\neq 0$, $\Phi_M$ is given by 
\be
\ba
&{ \sqrt 2\over |z_{+}|}
\sum_{h\ge 0}
\sum_{h^+,h^-}
{h!(-z_{+}^2/\pi)^h\over (2i)^{h^++h^-}}{h^+\choose h}{h^-\choose h}
\sum_{j}
\sum_{\lambda_K\in K}
{(-\Delta)^j(\bar \phi_{h^+,h^-})(\tilde{P}(\lambda))\over (8\pi)^j j!}\\
&\cdot \sum_{l,t}q^{l(\lambda_K,(-z'+{z_+\over 2z_+^2}+{z_-\over 2z_-^2}))}c(\lambda_K^2,t)l^{h^++h^--2h}\left({l\over 2|z_+||\tilde{P}_+(\lambda_K)|}\right)^{h-h^+-h^--j-t+{b^+\over 2}+m^+-3/2}\\
&K_{h-h^+-h^--j-t-b^+/2+m^+-3/2}\left({2\pi l |\tilde{P_+}(\lambda_K)|\over |z_+|}\right).
\ea
\ee
For $\tilde{P}_+(\lambda)=0$, the last two factors have to be replaced by the analytic continuation at $\epsilon=0$ of 

\be
\ba
&\hspace{-2cm}\left({\pi l^2\over 2z_+^2}\right)^{h-h^+-h^--j-t+b^+/2+m^+-3/2-\epsilon}\\
&\cdot \Gamma(-h+h^++h^-+j+t-b^+/2-m^++3/2+\epsilon).
\ea
\ee
%
\section{Instanton tables and heterotic-type II duals}\label{ap:data}
Table \ref{tab:dualCY} lists the dual K3-fibrations for the $\mbb{Z}_N$-orbifolds defined in table \ref{tab:orb} \cite{afiq}.

\begin{table}[!h]
\begin{center}
\begin{tabular}{|c|c|c|c|}
\hline
Type & Group & $(n_h,n_v)$ &CY weights\\
\hline
\hline
$\mbb{Z}_2, 3+8\ {\rm WL}$&$SU(4)\times E_8'\times U(1)^4$&$(167,15)$&$(1,1,12,16,18,20)$\\
$\mbb{Z}_2, 2+8 \ {\rm WL}$&$SU(3)\times E_8'\times U(1)^4$&$(230,14)$&$(1,1,12,16,18)$\\
$\mbb{Z}_2, 1+8 \ {\rm WL}$&$SU(2)\times E_8'\times U(1)^4$&$(319,13)$&$(1,1,12,16,30)$\\
$\mbb{Z}_2, 0+8 \ {\rm WL}$&$E_8'\times U(1)^4$&$(492,12)$&$(1,1,12,28,42)$\\
\hline\hline
$\mbb{Z}_3, 3+6 \ {\rm WL}$&$SU(4)\times E_6'\times U(1)^4$&$(129,13)$&$(1,1,6,10,12,14)$\\
$\mbb{Z}_3, 2+6 \ {\rm WL}$&$SU(3)\times E_6'\times U(1)^4$&$(168,12)$&$(1,1,6,10,12)$\\
$\mbb{Z}_3, 1+6 \ {\rm WL}$&$SU(2)\times E_6'\times U(1)^4$&$(221,11)$&$(1,1,6,10,18)$\\
$\mbb{Z}_3, 0+6 \ {\rm WL}$&$E_6'\times U(1)^4$&$(322,10)$&$(1,1,6,16,24)$\\
\hline\hline
$\mbb{Z}_4, 3+4 \ {\rm WL}$&$SU(4)\times SO(8)'\times U(1)^4$&$(123,11)$&$(1,1,4,8,10,12)$\\
$\mbb{Z}_4, 2+4 \ {\rm WL}$&$SU(3)\times SO(8)'\times U(1)^4$&$(154,10)$&$(1,1,4,8,10)$\\
$\mbb{Z}_4, 1+4 \ {\rm WL}$&$SU(2)\times SO(8)'\times U(1)^4$&$(195,9)$&$(1,1,4,8,14)$\\
$\mbb{Z}_4, 0+4 \ {\rm WL}$&$SO(8)'\times U(1)^4$&$(272,8)$&$(1,1,4,12,18)$\\
\hline\hline
$\mbb{Z}_6, 3+0 \ {\rm WL}$&$SU(4)\times E_6'\times U(1)^4$&$(139,7)$&$(1,1,2,6,8,10)$\\
$\mbb{Z}_6, 2+0 \ {\rm WL}$&$SU(3)\times E_6'\times U(1)^4$&$(162,6)$&$(1,1,2,6,8)$\\
$\mbb{Z}_6, 1+0 \ {\rm WL}$&$SU(2)\times E_6'\times U(1)^4$&$(191,5)$&$(1,1,2,6,10)$\\
$\mbb{Z}_6, 0+0 \ {\rm WL}$&$E_6'\times U(1)^4$&$(244,4)$&$(1,1,2,8,12)$\\
\hline
\end{tabular}
\caption{The chains of heterotic-type II duals studied in \cite{afiq}}\label{tab:dualCY}
\end{center}
\end{table}
Tables \ref{tab:first}--\ref{tab:last} give instanton numbers at $g=0,\cdots 4$ for the $\mbb{Z}_{2,3,4,6}$ orbifolds defined in table \ref{tab:orb}. 
\begin{table}[!h]
\begin{center}\begin{tabular}{|r|r|r|r|r|r|r|r|r|}
\hline
$g$& ${p^2\over 2}=-1$&$0$&$1$&$2$&$3$&$4$&$5$&$6$\\
\hline
0&-2&960&56808&1364480&20920140&240357888&2244734960&17884219392\\
\hline
1&0&4&-1908&-119360&-3077460&-50495040 &-617959240 &-6118785792\\
\hline
2&0&0&-6&2848&185694&5045376&87240260&1122823296\\
\hline
3&0&0&0&8&-3780&-255792 &-7276660 &-131766240\\
\hline
4&0&0&0&0&-10&4704&329630&9782592\\
\hline
\end{tabular}
\caption{$\mbb{Z}_2$, 8 Wilson lines, dual to $X^{1,1,12,28,42}$}\label{tab:first}
\vspace{.4cm}
\begin{tabular}{|r|r|r|r|r|r|r|r|r|r|r|}
\hline
$g$& ${p^2\over 2}=-1$&$-{1\over 4}$&$0$&${3\over 4}$&1&${7\over 4}$&$2$&${11\over 4}$&3&${15\over 4}$\\
\hline
0&-2&176&612&12672&30240&320976&661696&5031040&9509328&58372272\\
\hline
1&0&0&4&-352&-1212 &-26400 &-64136 &-719392 &-1509700 &-12091776 \\
\hline
2&0&0&0&0&-6&528&1804&40832&100422&1173600\\
\hline
3&0&0&0&0&0&0&8&-704&-2388&-55968\\
\hline
4&0&0&0&0&0&0&0&0&-10&880\\
\hline
\end{tabular}
\caption{$\mbb{Z}_2$, 8+1 Wilson lines, dual to $X^{1,1,12,16,30}$}
\vspace{.4cm}
\begin{tabular}{|r|r|r|r|r|r|r|r|r|r|r|}
\hline
$g$& ${p^2\over 2}=-1$&$-{1\over 3}$&$0$&${2\over 3}$&$1$&${5\over 3}$&$2$&${8\over 3}$&$3$&${11\over 3}$\\
\hline
0&-2&90&432&5904&18252&142146&365600&2144016&4936140&24107760\\
\hline
1&0&0&4&-180 &-852 &-12348 &-39080 &-320436 &-844140&-5189400\\
\hline
2&0&0&0&0&-6&270&1264&19152&61578&524952\\
\hline
3&0&0&0&0&0&0&8&-360&-1668&-26316\\
\hline
4&0&0&0&0&0&0&0&0&-10&450\\
\hline
\end{tabular}
\caption{$\mbb{Z}_2$, 8+2 Wilson lines, dual to $X^{1,1,12,16,30}$}
\vspace{.4cm}
\begin{tabular}{|r|r|r|r|r|r|r|r|r|r|r|r|r|}
\hline
$g$& ${p^2\over 2}=-1$&$-{1\over 2}$&$-{3\over 8}$&$0$&${1\over 2}$&${5\over 8}$&$1$&${3\over 2}$&${13\over 8}$&$2$&${5\over 2}$\\
\hline
0&-2&28&64&304&2144&3392&11412&52144&75136&211040&781312\\
\hline
1&0&0&0&4&-56&-128&-596&-4456&-7168&-24632 &-117376\\
\hline
2&0&0&0&0&0&0&-6&84&192&880 &6880 \\
\hline
3&0&0&0&0&0&0&0&0&0&8&-112\\
\hline
\end{tabular}
\caption{$\mbb{Z}_2$, 8+3 Wilson lines, dual to $X^{1,1,12,16,18}$}
\vspace{.4cm}
\begin{tabular}{|r|r|r|r|r|r|r|r|r|r|r|r|r|}
\hline
$g$& ${p^2\over 2}=-1$&$-{3\over 5}$&$-{2\over 5}$&$0$&${2\over 5}$&${3\over 5}$&$1$&${7\over 5}$&${8\over 5}$&$2$&${12\over 5}$\\
\hline
0&-2&14&52&200&1020&2158&7068&23916&43080&122840&347376\\
\hline
1&0&0&0&4&-28&-104&-388&-2124&-4628 &-15320 &-54064\\
\hline
2&0&0&0&0&0&0&-6&42&156&568 &3284 \\
\hline
3&0&0&0&0&0&0&0&0&0&8&-56\\
\hline
\end{tabular}
\caption{$\mbb{Z}_2$, 8+4 Wilson lines, dual to $X^{1,1,12,16,18,20}$}
\vspace{.5cm}
\begin{tabular}{|r|r|r|r|r|r|r|r|r|r|r|r|r|}
\hline
$g$& ${p^2\over 2}=-1$&$-{1\over 2}$&$-{3\over 8}$&$0$&${1\over 2}$&${5\over 8}$&$1$&${3\over 2}$&${13\over 8}$&$2$\\
\hline
0&-2&8&24&264&9104&17272&86292&634464&1009936&3647120\\
\hline
1&0&0&0&4&-16 &-48 &-516 &-18256 &-34688 &-174152\\
\hline
2&0&0&0&0&0&0&-6&72&760&27440 \\
\hline
3&0&0&0&0&0&0&0&0&0&8\\
\hline
\end{tabular}
\caption{$\mbb{Z}_6$, 3 Wilson lines, dual to $X^{1,1,2,6,8,10}$}
\end{center}
\end{table}
\begin{table}
\begin{center}
\begin{tabular}{|r|r|r|r|r|r|r|r|r|r|r|r|r|}
\hline
$g$& ${p^2\over 2}=-1$&$0$&$1$&$2$&$3$&$4$&$5$&$6$\\
\hline
0&-2&624&54792&1609088&28265184&360251424&3659578208&31296575232\\
\hline
1&0&4 &-1236 &-113312&-3551892&-66631944&-903741184&-9729986112\\
\hline
2&0&0&-6&1840&174270&5731824&113066144&1610777952\\
\hline
3&0&0&0&8&-2436&-237648 &-8154292&-168125136\\
\hline
4&0&0&0&0&-10&3024& 303422&10826544\\
\hline
\end{tabular}
\caption{$\mbb{Z}_3$, 6 Wilson lines, dual to $X^{1,1,6,16,24}$}
\vspace{.5cm}
\begin{tabular}{|r|r|r|r|r|r|r|r|r|r|r|r|r|}
\hline
$g$& ${p^2\over 2}=-1$&$-{1\over 4}$&$0$&${3\over 4}$&$1$&${7\over 4}$&$2$&${11\over 4}$&$3$\\
\hline
0&-2&104&420&11856&30240 &373464 &801472 & 6750016&13138500\\
\hline
1&0&0&4 & -208&-828&-24336 &-62984 &-818896 &-1787716\\
\hline
2&0&0&0&0& -6&312 &1228 & 37232& 97350\\
\hline
3&0&0&0&0&0&0&8&-416&-1620\\
\hline
\end{tabular}
\caption{$\mbb{Z}_3$, 6+1 Wilson lines, dual to $X^{1,1,6,10,18}$}
\vspace{.4cm}
\begin{tabular}{|r|r|r|r|r|r|r|r|r|r|r|r|r|}
\hline
$g$& ${p^2\over 2}=-1$&$-{1\over 3}$&$0$&${2\over 3}$&$1$&${5\over 3}$&$2$&${8\over 3}$&$3$\\
\hline
0&-2&54&312& 5616& 18900& 167778& 454688&2914704&6972912\\
\hline
1&0&0&4 &-108&-612&-11556&-39656&-369684&-1025244\\
\hline
2&0&0&0&0& -6&162 & 904& 17712&61602\\
\hline
3&0&0&0&0&0&0&8&-216&-1188\\
\hline
\end{tabular}
\caption{$\mbb{Z}_3$, 6+2 Wilson lines, dual to $X^{1,1,6,10,12}$}
\vspace{.5cm}
\begin{tabular}{|r|r|r|r|r|r|r|r|r|r|r|r|r|}
\hline
$g$& ${p^2\over 2}=-1$&$-{1\over 2}$&$-{3\over 8}$&$0$&${1\over 2}$&${5\over 8}$&$1$&${3\over 2}$&${13\over 8}$&$2$&${5\over 2}$\\
\hline
0&-2&16& 40& 232& 2024&3320 &12228&61600 &90592& 269456 &1065784\\
\hline
1&0&0&0 &4&-32&-80&-452&-4144&-6880&-25832&-135472\\
\hline
2&0&0&0&0& 0&0 & -6&48&120&664&6328\\
\hline
3&0&0&0&0&0&0&0&0&0&8&-64\\
\hline
\end{tabular}
\caption{$\mbb{Z}_3$, 6+3 Wilson lines, dual to $X^{1,1,6,10,12}$}
\vspace{.4cm}
\begin{tabular}{|r|r|r|r|r|r|r|r|r|r|r|r|r|}
\hline
$g$& ${p^2\over 2}=-1$&$0$&$1$&$2$&$3$&$4$&$5$\\
\hline
0&-2&528&90036&3679520&80559180&1212246784&14073864648\\
\hline
1&0&4 &-1044 &-183224&-7903452&-183923136&-2938551600\\
\hline
2&0&0&-6&1552&278466&12502704&304651808\\
\hline
3&0&0&0&8&-2052 &-375744 &-17481820 \\
\hline
4&0&0&0&0&-10&2544&475034\\
\hline
\end{tabular}
\caption{$\mbb{Z}_4$, 4 Wilson lines, dual to $X^{1,1,4,12,18}$}
\vspace{.5cm}
\begin{tabular}{|r|r|r|r|r|r|r|r|r|r|r|r|r|}
\hline
$g$& ${p^2\over 2}=-1$&$-{1\over 4}$&$0$&${3\over 4}$&$1$&${7\over 4}$&$2$&${11\over 4}$&$3$\\
\hline
0&-2&80&372&18432&52428&832848&1908808&18982912&38738880\\
\hline
1&0&0&4 &-160&-732 &-37344 &-107072 &-1776928&-4135132\\
\hline
2&0&0&0&0& -6&240&1084&56576&163146\\
\hline
3&0&0&0&0&0&0&8&-320&-1428\\
\hline
\end{tabular}
\caption{$\mbb{Z}_4$, 4+1 Wilson lines, dual to $X^{1,1,4,8,14}$}
\end{center}
\end{table}
\begin{table}[!h]
\begin{center}
\begin{tabular}{|r|r|r|r|r|r|r|r|r|r|r|r|r|}
\hline
$g$& ${p^2\over 2}=-1$&$-{1\over 3}$&$0$&${2\over 3}$&$1$&${5\over 3}$&$2$&${8\over 3}$&$3$\\
\hline
0&-2& 42& 288& 8928 & 34488 & 381894 & 1127168& 8355360 & 21263796\\
\hline
1&0&0&4 &-84&-564&-18108&-70688&-817692&-2463540\\
\hline
2&0&0&0&0& -6& 126& 832&27456&107982\\
\hline
3&0&0&0&0&0&0&8&-168&-1092\\
\hline
\end{tabular}
\caption{$\mbb{Z}_4$, 4+2 Wilson lines, dual to $X^{1,1,4,8,10}$}
\vspace{.5cm}
\begin{tabular}{|r|r|r|r|r|r|r|r|r|r|r|r|r|}
\hline
$g$& ${p^2\over 2}=-1$&$-{1\over 2}$&$-{3\over 8}$&$0$&${1\over 2}$&${5\over 8}$&$1$&${3\over 2}$&${13\over 8}$&$2$&${5\over 2}$\\
\hline
0&-2& 12& 32& 224 & 3136 & 5536 &23392 &139688 &213248 &694400 & 3063424\\
\hline
1&0&0&0 &4&-24&-64&-436&-6344&-11264&-48112&-298288\\
\hline
2&0&0&0&0& 0&0 & -6& 36&96&640&9600\\
\hline
3&0&0&0&0&0&0&0&0&0&8&-48\\
\hline
\end{tabular}
\caption{$\mbb{Z}_4$, 4+3 Wilson lines, dual to $X^{1,1,4,8,10,12}$}\label{tab:last}
\end{center}
\end{table}

\clearpage

\bibliographystyle{fullsort}
\providecommand{\href}[2]{#2}\begingroup\raggedright\endgroup
\end{document}